\documentclass[apj]{emulateapj}
 
\newcommand{\kms}       {\mbox{km s$^{-1}$}}%
\newcommand{\kmsMpc}    {\mbox{km s$^{-1}$ Mpc$^{-1}$}}%
\newcommand{\magarc}    {\ensuremath{\mbox{mag arcsec}^{-2}}}%

\newcommand{\msun}      {\mbox{$M_\odot$}} 

\shortauthors{Kannappan, Guie, \& Baker}
\shorttitle{Blue-Sequence E/S0 Galaxies}
\slugcomment{submitted}
 
\begin{document}

\title{E/S0 Galaxies on the Blue Color-Stellar Mass Sequence at
  $z=0$:\\ Fading Mergers or Future Spirals?}
\author{Sheila J. Kannappan\altaffilmark{1},
	Jocelly M. Guie\altaffilmark{2},
	and Andrew J. Baker\altaffilmark{3}}
\altaffiltext{1}{Department of Physics and Astronomy, 
  University of North Carolina, 290 Phillips Hall CB 3255, 
  Chapel Hill, NC 27599; sheila@physics.unc.edu}
\altaffiltext{2}{Department of Astronomy, The University of Texas at
        Austin, 1 University Station C1400, Austin, TX 78712-0259;
        jocelly@mail.utexas.edu}
\altaffiltext{3}{Department of Physics and
  Astronomy, Rutgers, the State University of New Jersey, 136
  Frelinghuysen Road, Piscataway, NJ 08854-8019;
  ajbaker@physics.rutgers.edu}

\begin{abstract}
We identify a population of morphologically defined E/S0 galaxies
lying on the locus of late-type galaxies in color--stellar mass space
--- the ``blue sequence'' --- at the present epoch.  Using three
samples (from the Nearby Field Galaxy Survey or NFGS, a merged
HyperLeda/SDSS/2MASS catalog, and the NYU Value Added Galaxy Catalog),
we analyze blue-sequence E/S0s with stellar masses $\ga$10$^8$ \msun,
arguing that individual objects may be evolving either up toward the
red sequence or down into the blue sequence.  Blue-sequence E/S0
galaxies become more common with decreasing stellar mass, comprising
$\la$2\% of E/S0s near the ``shutdown mass'' $M_s$ $\sim$ 1--2
$\times$ 10$^{11}$ \msun, increasing to $\ga$5\% near the ``bimodality
mass'' $M_b$ $\sim$ 3 $\times$ 10$^{10}$ \msun, and sharply rising to
$\ga$20--30\% below the ``threshold mass'' $M_t$ $\sim$ 4--6 $\times$
10$^{9}$ \msun, down to our completeness analysis limit at
$\sim$10$^9$ \msun.  The strong emergence of blue-sequence E/S0s below
$M_t$ coincides with a previously reported global increase in mean
atomic gas fractions below $M_t$ for galaxies of all types on both
sequences, suggesting that the availability of cold gas may be basic
to blue-sequence E/S0s' existence.  Environmental analysis reveals
that many sub-$M_b$ blue-sequence E/S0s reside in low to intermediate
density environments.  Thus the bulk of the population we analyze
appears distinct from the generally lower-mass cluster dE population;
S0 morphologies with a range of bulge sizes are typical.  In
mass-radius and mass-$\sigma$ scaling relations, blue-sequence E/S0s
are more similar to red-sequence E/S0s than to late-type galaxies, but
they represent a transitional class.  While some of them, especially
in the high-mass range from $M_b$ to $M_s$, resemble major-merger
remnants that will likely fade onto the red sequence, most
blue-sequence E/S0s below $M_b$ show signs of disk and/or pseudobulge
building, which may be enhanced by companion interactions.  The blue
overall colors of blue-sequence E/S0s are most clearly linked to blue
outer disks, but also reflect blue centers and more frequent
blue-centered color gradients than seen in red-sequence E/S0s.
Notably, all E/S0s in the NFGS with polar or counterrotating gas lie
on or near the blue sequence, and most of these systems show signs of
secondary stellar disks forming in the decoupled gas.  From star
formation rates and gas fractions, we infer significant recent and
ongoing morphological transformation in the blue-sequence E/S0
population, especially below $M_b$.  We argue that sub-$M_b$
blue-sequence E/S0s occupy a ``sweet spot'' in stellar mass and
concentration, with both abundant gas and optimally efficient star
formation, which may enable the formation of large spiral disks.  Our
results provide evidence for the importance of disk rebuilding after
mergers, as predicted by hierarchical models of galaxy formation.
\end{abstract}


\section{Introduction}
\label{sec:intro}

Modern galaxy surveys find that most galaxies occupy two distinct loci
in color-stellar mass space, the ``red sequence'' and the ``blue
sequence'' \citep[or ``cloud'' or ``distribution,''
  etc.;][]{strateva.ivezi-c.ea:color,baldry.glazebrook.ea:quantifying,bell.wolf.ea:nearly}.
Because the high-mass end of the red sequence corresponds to the
well-known color-magnitude relation of cluster E/S0 galaxies, it is
natural to identify red-sequence galaxies with E/S0 types and
blue-sequence galaxies with spiral/irregular types, except for some
contamination of the red sequence by dust-reddened late-type systems.

Here we show that this basic morphology--color correspondence begins
to fail for stellar masses $M_*$ $\la$ 1--2 $\times$ $10^{11}$ \msun,
as morphologically defined E/S0 galaxies start to appear on the blue
sequence in $z=0$ surveys \citep[see
also][]{bamford.nichol.ea:galaxy}.  The failure becomes dramatic below
a threshold mass $M_t$ of 4--6 $\times$ 10$^9$ \msun, matching the
mass scale below which the mean atomic gas content of low-$z$ galaxies
increases substantially on both sequences
\citep[][]{kannappan:linking,kannappan.wei:galaxy}.
\footnote{\citet{kannappan:linking} incorrectly associated the
  gas-richness threshold mass with the 3 $\times$ 10$^{10}$ \msun\
  ``bimodality'' mass scale of \citet{kauffmann.heckman.ea:dependence}
  based on an assumed correspondence between the stellar mass
  calibrations of Kauffmann et al. and
  \citet{bell.mcintosh.ea:optical}. On further investigation (see both
  \S~\ref{sec:massmethods} and \citealt{kannappan.wei:galaxy}), the
  gas-richness threshold mass in fact corresponds to $\sim$4--6
  $\times$ 10$^9$ \msun\ and marks the lower edge of a transitional
  mass range centered on the bimodality mass.}

A specific stellar mass scale linked to shifts in galaxy properties
was first highlighted by \citet{kauffmann.heckman.ea:dependence}, who
found that galaxy star formation histories (SFHs) are qualitatively
different on either side of a crossover or bimodality mass $M_b$
$\sim$ $3\times10^{10}$ \msun.  Their analysis shows the transitional
mass range starting at $\log{M_*}/\msun$ $\sim$ 9.5--10, below which
galaxies have bursty SFHs, and ending at $\log{M_*/\msun}$ $\sim$
11--11.5, beyond which SFHs reflect uniformly ancient stellar
populations.  Structural properties also change, with high-mass
galaxies having higher concentration, higher surface brightness, and a
higher ratio of spheroids to disks
\citep{kauffmann.heckman.ea:dependence,driver.allen.ea:millennium}.
In a more focused study of edge-on disk galaxies,
\citet{dalcanton.yoachim.ea:formation} report the emergence of bulges
as well as efficient star formation, as evidenced by prominent dust
lanes, above $V_c$ $\sim$ 120 \kms\ (corresponding to $\log{M_*/\msun}
\sim 9.5$ in the stellar-mass Tully-Fisher relation, see
\S~\ref{sec:scaling} below).  \citet{baldry.glazebrook.ea:quantifying}
tie mass-dependent shifts in galaxy properties to the red and blue
sequences by showing that the red sequence becomes numerically
dominant above $M_*$ $\sim$ 2--3 $\times$ 10$^{10}$ \msun.

Just about every idea advanced to explain why galaxy properties change
at specific mass scales involves gas physics.  Abrupt shifts in star
formation efficiency and gas richness near $M_t$ have been linked to
the interplay of gas infall, supernova-driven winds, and changes in
surface mass density
\citep{dalcanton.yoachim.ea:formation,dalcanton:metallicity}.  The
buildup of the red sequence above $M_b$ has been linked to
``quenching'' processes that shut down star formation, such as gas
loss caused by AGN feedback, consumption of all available gas in the
wake of violent merging, ram-pressure
stripping/harassment/strangulation in clusters, and/or shock-heating
that turns off efficient ``cold-mode'' gas accretion
\citep[][]{baldry.glazebrook.ea:quantifying,kenney..ea:vla,springel.di-matteo.ea:black,dekel.birnboim:galaxy,cattaneo.dekel.ea:modelling,faber.willmer.ea:galaxy}.
Some combination of these processes may be required to ensure that
quenching is efficient, permanent, and mass-dependent
\citep[][]{dekel.birnboim:galaxy}.

Higher-redshift studies now suggest that the crossover mass between
disks and spheroids has evolved downward over time from $M_*$ $\sim$
1--$2 \times10^{11}$ \msun\ at $z\sim1$ --- notably, equal to the
shutdown mass $M_s$ marking the upper end of the transitional mass
range today --- to today's crossover mass near $M_b$
(\citealt{bundy.ellis.ea:mass*1};
\citealt{franceschini.rodighiero.ea:cosmic}; see also
\citealt{cimatti.daddi.ea:mass}).  This trend is coincident with many
other signs of ``downsizing'' \citep[][]{cowie.songaila.ea:new} in the
galaxy formation process, including the downwardly evolving upper mass
threshold for strong star formation \citep[e.g.,][]
{juneau.glazebrook.ea:cosmic,kodama.yamada.ea:down-sizing,tresse.ilbert.ea:cosmic}
as well as a similarly evolving upper threshold for irregular
morphologies, abundant blue compact galaxies, and strong kinematic
disturbances
\citep{mall-en-ornelas.lilly.ea:internal,kannappan.barton:tools}.
Whether downsizing can be traced back to still higher redshifts is
unclear, as the red sequence is only tentatively detected beyond
$z\sim1.5$ (comparing \citealt{kriek..ea:detection} and
\citealt{cirasuolo.mclure.ea:evolution}).  Semianalytic models that
include cold accretion predict that the highest mass galaxies,
residing in $\sim$ 10$^{12}$ \msun\ dark matter halos (implying
stellar masses near $M_s$), formed rapidly at very high redshift
before developing stable shocks that inhibited further star formation
\citep{cattaneo.dekel.ea:modelling}.

From these results, we infer the importance of three mass regimes for
our $z=0$ analysis: (1) near the shutdown mass $M_s$, at which
blue-sequence E/S0s first emerge, and above which nearly all galaxies
are old and red; (2) between $M_s$ and $M_t$, where the signatures of
downsizing from $z=1-0$ may remain; and (3) below the threshold mass
$M_t$, where blue-sequence E/S0s, bursty SFHs, and high gas fractions
become suddenly abundant. The surveys considered here include few
galaxies below $M_*$ $\sim$ a few $\times$ 10$^8$ \msun, so we largely
omit the abundant cluster dE population that dominates the faint end
of the red sequence.  In fact, many E/S0s between 10$^8$--10$^{10.5}$
\msun\ occupy low-to-intermediate density environments
(\S~\ref{sec:massenvt}; \citealt{hogg.blanton.ea:overdensities};
\citealt{trentham.sampson.ea:galaxy}).

Given the close association of blue-sequence E/S0s with mass regimes
characterized by active or recent evolution, their intriguing
mismatched color-morphology status could indicate a transitional state
in color (e.g., a fading starburst galaxy after a merger), morphology
(e.g., a disk-building system), or both (e.g, a gas-rich merger remnant
regenerating a young disk from tidal debris).  Put differently,
blue-sequence E/S0s may be evolving from blue to red, from red to
blue, or wholly within the blue sequence.  Or they may not be evolving
much at all.

In this work we document the blue-sequence E/S0 population
systematically for the first time, exploring properties and
demographics to constrain the disk-building and fading-starburst
scenarios.  We ask the reader to suspend all preconceptions about E/S0
colors, gas contents, masses, environments, and even bulge-to-disk
ratios, as many accepted ``facts'' about E/S0s reflect the properties
of the high-mass red sequence.  Section~\ref{sec:data} describes our
survey samples, our morphological definition of an E/S0, how we
measure stellar mass, and how we divide the red and blue sequences.
Section~\ref{sec:exist} provides a demographic overview of the
blue-sequence E/S0 population, including frequency among all E/S0s,
distribution by mass and environment, and dynamical scaling relations.
Section~\ref{sec:props} examines the detailed characteristics of
blue-sequence E/S0s, including morphological substructure, gas content
and star formation, and clues to interaction status.
Section~\ref{sec:peckin} follows up on an intriguing association of
gas-stellar counterrotation and polar rings with blue-sequence E/S0s,
providing an in-depth analysis of new stellar kinematic data for five
such galaxies to probe the possibility of secondary stellar disk
growth in these systems.  Finally, Section~\ref{sec:evol} reviews our
results in the context of different evolutionary scenarios and
discusses implications for disk formation from $z=1-0$.

\section{Data and Methods}
\label{sec:data}

We assume $H_0=70$ \kmsMpc and $d=cz/H_0$ throughout.  At the low
redshifts of our sample galaxies, neglecting $\Lambda$ introduces
errors $\la$0.02 dex in stellar mass.

\subsection{Samples}
\label{sec:samples}

\begin{figure*}[t]
\plottwo{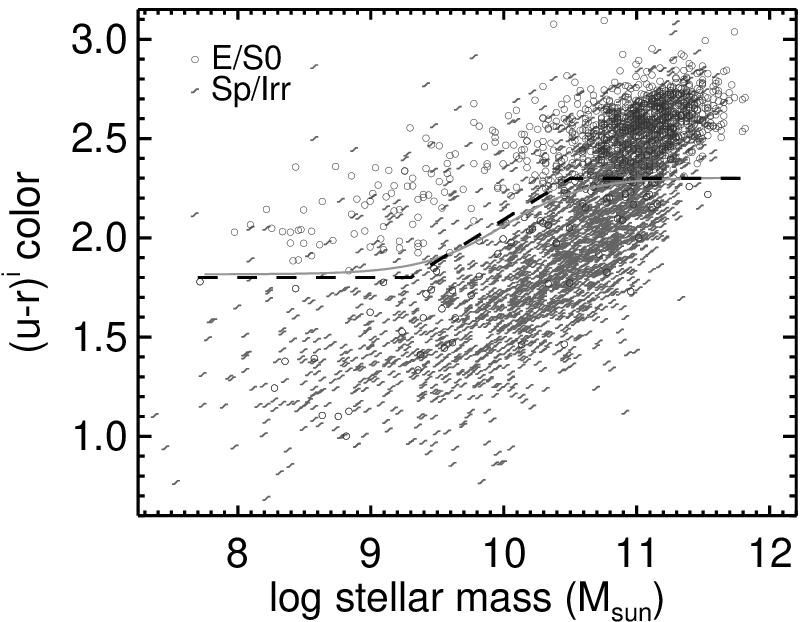}{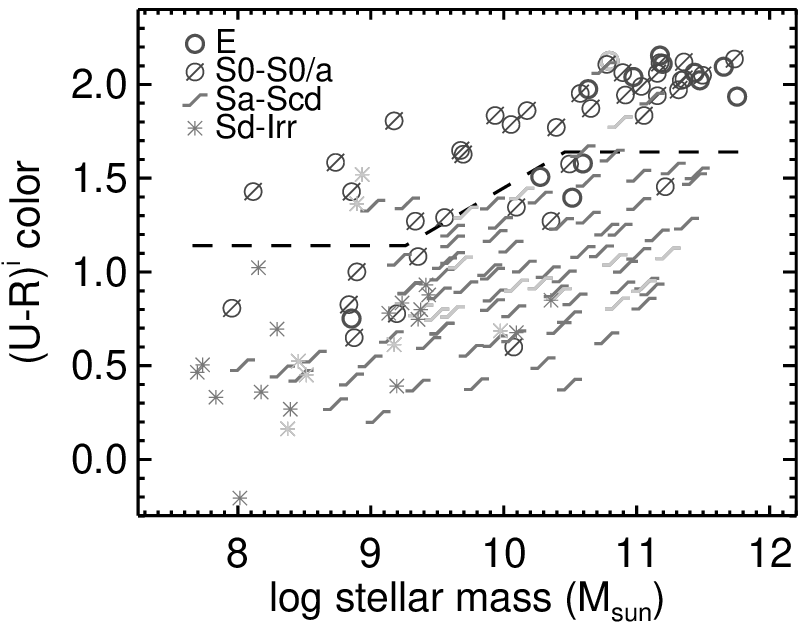}
\caption{Distribution of morphologies in color vs.\ stellar mass
  space, for {\it (a)} the HyperLeda+ sample and {\it (b)} the NFGS
  sample.  In both panels, the dashed line divides the red sequence
  from the blue sequence, with a calibrated offset from $u-r$ to $U-R$
  (\S~\ref{sec:divider}) and a shift of 0.04 dex between masses
  determined with SDSS vs.\ NFGS photometry
  (\S~\ref{sec:massmethods}).  The gray curve in the panel {\it a} is
  the sequence divider of \citet{baldry.glazebrook.ea:quantifying},
  approximately converted from $L_r$ to $M_*$ units.  Blue-sequence
  E/S0s in the HyperLeda+ sample have been vetted as described in
  \S~\ref{sec:samples}, but late types and red-sequence E/S0s in the
  HyperLeda+ sample have not been reclassified; thus the blue-sequence
  E/S0s shown in panel {\it a} represent a lower limit on the
  fractional size of the population. Lighter gray symbols in panel
  {\it b} show galaxies with visually noted defects in 2MASS
  photometry; see footnote~\ref{fn:k}.}
\label{fg:cmtboth}
\end{figure*}

We analyze two primary samples, one drawn from the Nearby Field Galaxy
Survey \citep[NFGS,][ hereafter J00]{jansen.franx.ea:surface} and the
other drawn from the HyperLeda, Sloan Digital Sky Survey (SDSS) DR4,
and Two Micron All-Sky Survey (2MASS) databases \citep[hereafter,
  ``HyperLeda+'' sample;][]{paturel.petit.ea:hyperleda,
  adelman-mccarthy.agueros.ea:fourth,jarrett.chester.ea:2mass}.  These
samples are shown in Fig.~\ref{fg:cmtboth}.  The NFGS sample offers a
homogeneous and high-quality data set while the HyperLeda+ sample
offers the statistical advantage of thousands of galaxies.  A third
sample, the NYU Value-Added Galaxy Catalog low-redshift sample
\citep[VAGC,][]{blanton.schlegel.ea:new}, is used to evaluate
completeness in the HyperLeda+ sample.  We describe the samples here
and amplify on the morphological criteria used to identify E/S0s in
\S~\ref{sec:etgdef}.

\subsubsection{The Nearby Field Galaxy Survey (NFGS) Sample}

The NFGS database includes $\sim$200 galaxies selected to fairly
represent the natural abundance of morphologies in the local universe
over a wide range in luminosity.  Notwithstanding its name, the survey
spans a variety of environments including clusters.  NFGS galaxies
were given numerical morphological types as part of the CfA~1 redshift
survey \citep{huchra.davis.ea:survey} and later reclassified by J00 on
the same deVaucoleurs-based system, using new high-quality CCD
imaging.\footnote{We convert the type Pec galaxy UGC\,9562, a polar
ring galaxy with an S0 host, to type S0 for this paper.  We also
reclassify UGC\,6570 (=A11332+3536) from S0 to S0/a following
\citet{kannappan.jansen.ea:forming}.  In all other cases we use the
types given by J00 for the NFGS.}  Because the NFGS was selected
without explicit bias in color or morphology, the properties of
blue-sequence E/S0s in this survey should be reasonably representative,
within the constraints of $B$-band selection and small number 
statistics.

The NFGS sample allows us to examine a broad sample of E/S0s with
supporting data not available for the HyperLeda+ sample, most notably,
integrated spectrophotometry
\citep{jansen.fabricant.ea:spectrophotometry} and gas and stellar
kinematics \citep[][see also new data in
\S~\ref{sec:cr}]{kannappan.fabricant:broad,kannappan.fabricant.ea:physical,kannappan.fabricant.ea:kinematics}. Other
data used in analyzing the NFGS sample include $UBR$ photometry and
Virgocentric-infall corrected redshifts from J00, 2MASS $K$-band
photometry from \citet[][]{jarrett.chester.ea:2mass}, and atomic gas
masses derived from the homogenized HI catalog of
\citet{paturel.theureau.ea:hyperleda} using $M_{\rm
HI+He}=1.4(2.36\times10^5 fd^2)$ \msun, where $f$ is the integrated
line flux in Jy\, km\, s$^{-1}$, $d$ is the distance in Mpc, and the
factor of 1.4 represents the helium mass correction. SDSS $ugriz$ data
are available for $\sim$60\% of the NFGS; we use these magnitudes to
calibrate shifts in colors and mass estimates between the $UBRJHK$
(primary NFGS) and $ugrizJHK$ (HyperLeda+) systems.

Our primary NFGS sample consists of 164 galaxies, including 52 E/S0s.
We reject galaxies with missing spectra or point-source morphologies.
We also exclude a possible blue-sequence E/S0, NGC\,2824, with severe
contamination from a nearby bright star; however, we retain one
red-sequence E/S0, NGC\,3605, whose $B$ and $R$-band magnitudes are
mildly affected by blending with a companion, but whose $U-R$ color
can still be estimated reasonably well based on stellar population
fits to the full SED.  All sources in this primary sample have
$K$-band fluxes with error $<$0.3 mag.\footnote{\label{fn:k} Detailed
  inspection reveals galaxies for which 2MASS data are available but
  pipeline data products are defective, due to poor deblending,
  electronic glitches, cut-off edges, or other technical problems
  \citep{kannappan.fabricant.ea:kinematics}. This flagging would
  reject some galaxies that pass our $K$-band error criterion, but we
  allow these galaxies in this paper to facilitate uniform comparison
  with the HyperLeda+ and NYU VAGC samples. Fig.~\ref{fg:cmtboth}b
  indicates galaxies that pass the $K$-band criterion but not the more
  detailed inspection in a lighter shade of gray; all but one are
  late-type galaxies.}

Both spectra and photometry are corrected for foreground extinction
using the maps of \citet{schlegel.finkbeiner.ea:maps} and, for the
spectra, the Milky Way extinction curve of
\citet{odonnell:rnu-dependent} as given by \citet{mccall:on}.  We use
total, extrapolated magnitudes for stellar mass estimation (along with
spectra, \S~\ref{sec:massmethods}).  The red and blue sequences are
divided using isophotal $U-R$ colors measured within the $B$-band 26
\magarc\ isophote, to minimize extrapolation errors.  We estimate
$k$-corrections from the SED fits used to determine stellar masses
(see \S~\ref{sec:massmethods}).  The notation $(U-R)^i$ signifies that
we also correct for internal extinction in a population-averaged way,
by applying the inclination-based corrections of
\citet{tully.pierce.ea:global} adapted to the NFGS and SDSS passbands
\citep[see][]{kannappan.fabricant.ea:physical,kannappan.fabricant.ea:kinematics}.
These corrections are appropriate for galaxies with gas; for galaxies
lacking extended emission lines, we assume zero internal extinction.

\subsubsection{The HyperLeda/SDSS/2MASS (HyperLeda+) and NYU VAGC Samples}

The HyperLeda+ sample is a larger, less uniform data set consisting of
3783 galaxies with morphological types listed in HyperLeda plus $ugr$
and $K$ magnitudes available from the SDSS and 2MASS galaxy catalogs
(within a 6$\arcsec$ matching radius and with magnitude errors $<0.3$
in $K$ and $<0.15$ in $ugr$).  We include only galaxies with diameter
$D_{25}$ less than 120$\arcsec$ to minimize pipeline reduction errors
in 2MASS and SDSS.  To improve classification reliability, we exclude
galaxies listed as ``multiple'' in HyperLeda, and we require diameter
$\geq$40$\arcsec$ and axial ratio $<$0.45 (inclination less than
$\sim$72$\degr$).  We adopt Virgocentric-infall corrected redshifts
from HyperLeda when available and otherwise use SDSS redshifts, which
we correct to the Local Group frame of reference.  Data for a handful
of catalog galaxies that pass our diameter cuts but have nominal
redshifts $cz>15000$ \kms\ are rejected as spurious.  We further
require $cz>500$ \kms\ and exclude sources with $cz<1500$ \kms\ if
they lack Virgocentric infall corrections.

For our working HyperLeda+ sample we have reclassified all candidate
blue-sequence E/S0 galaxies using SDSS $g$-band imaging, with
reclassifications homogenized between two independent observers (SJK
and AJB), taking the NFGS sample as a reference.  Starting with
HyperLeda numerical types $\leq$0 (S0/a), our reclassification
discards $\sim$60\% of candidate blue-sequence E/S0s as actually having
type Sa or later.  The remaining confirmed blue-sequence E/S0s
represent a lower limit to the true population, because we have not
retyped galaxies initially classified as late type to find mistyped
E/S0s.\footnote{ The HyperLeda+ sample used in this paper was defined
prior to a 2006 update to the HyperLeda database that corrected a
systematic error in galaxy radii and added new data.  Adopting the
revised radii would slightly change our existing sample, because two
galaxies would just miss the lower diameter limit, but would not
otherwise affect our analysis, as the radii we analyze come from the
2MASS catalog.  Using the revised database would also add and remove
some candidate E/S0s, because average numerical types can shift as new
data are added.  However, as all of our blue-sequence E/S0s are
independently verified by inspection of SDSS data, our data set
remains a robust minimum sampling of the blue-sequence E/S0 population.
We therefore retain the original sample.}

In analyzing the HyperLeda+ sample we employ total magnitudes
determined from extrapolated profile fits, i.e., SDSS model magnitudes
and 2MASS extrapolated magnitudes.  We determine $k$-corrections and
foreground extinction corrections as for the NFGS sample.  Internal
extinction corrections are also matched in principle.  However,
because we do not know which galaxies have extended ionized gas
emission lines (the criterion used to decide whether to apply an
internal extinction correction for an NFGS galaxy), we exploit a
division in mass--radius space that predicts surprisingly well which
galaxies in the NFGS have extended emission. Thus an internal
extinction correction is applied whenever the Petrosian $r$-band 90\%
light radius $r_{90}$ satisfies $\log{r_{90}/{\rm kpc}} > -4.55
+0.50\log{M_*/\msun}$.

We determine colors and masses for the NYU VAGC low-redshift sample
almost exactly as for the HyperLeda+ sample, to facilitate the
comparative analysis in \S~\ref{sec:massenvt}.  However, axial ratios
for VAGC galaxies are available only by cross-matching with the online
HyperLeda database, which yields values for only $\sim$60--70\% of the
VAGC galaxies we analyze (depending on sample definition,
\S~\ref{sec:massenvt}). Galaxies lacking axial ratio data are treated
as having no internal extinction.

\subsection{Definition of an E/S0}
\label{sec:etgdef}

Our definition of an E/S0 galaxy is {\it purely morphological,} based
on traditional by-eye classification of monochrome $B$ or $g$ band
images by multiple observers.  As discussed in \S~\ref{sec:samples},
we calibrate our reclassifications of candidate HyperLeda+
blue-sequence E/S0s against the classifications given for the NFGS in
J00, which were determined on a simplified version of the
deVaucoleurs-based system used for the CfA~1 redshift survey.
Numerical types $\leq0$ (S0/a) qualify as E/S0s, including all type E,
cE, S0, and S0/a galaxies.  We also include some peculiar galaxies
with predominantly spheroidal structure.  Our classification system
does not identify ``blue compact dwarfs'' (BCDs) as a distinct
category, but rather distributes them among early and late types
according to their primary morphology. Dwarf ellipticals (as the term
is used by \citealt{binggeli.sandage.ea:luminosity}) fall mostly below
our mass range, and their tendency to occur in clusters is
inconsistent with the available environmental data for our galaxies
(\S~\ref{sec:massenvt}).  However, we make no explicit exclusion based
on concentration or surface brightness, and we find a continuous range
of E/S0 properties.\footnote{It is worth recalling that the strong
dichotomy in dE vs. giant E properties found in the Kormendy Relation
and other scaling laws does not consider S0 galaxies, which are
typically either excluded or included only in their bulge component.
S0s display a broad range of intermediate morphologies and masses, and
S0s formed in mergers are likely to exhibit a wide variety of surface
brightness profiles as a function of progenitor gas richness.
Therefore an inclusive approach is physically motivated.}

\begin{figure*}[t]
\plotone{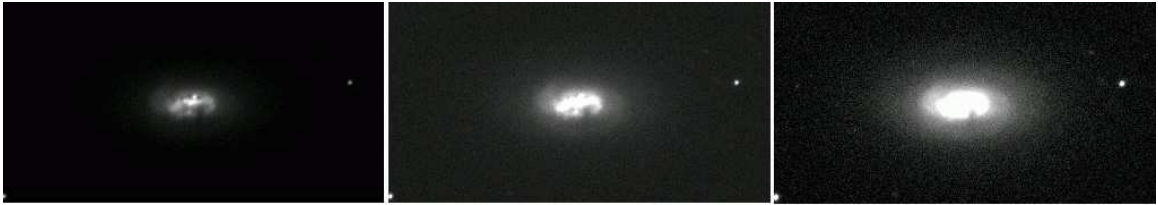}
\caption{Three $g$-band image stretches for a galaxy on the borderline
  between S0/a and Sa, UGC\,4902, which our merged classification
  lists as ``X,'' meaning literally too difficult to label either way.
  The finger of asymmetric dust to the south and the ambiguous,
  possibly one-sided arm structure make this a challenging galaxy to
  classify. Such borderline cases make up $\la$10\% of confirmed
  blue-sequence E/S0s in the HyperLeda+ sample.}
\label{fg:justone}
\end{figure*}

In practice, most E/S0s in the mass range of our analysis are S0 and
S0/a galaxies.  Our classification system downplays the importance of
bulge-to-disk ratio as a defining characteristic of S0s, requiring
only a bulge plus a smooth outer disk.  From this point of view, the
main criterion distinguishing an S0 from a spiral is the presence of
extended spiral structure in the latter \citep[as in the parallel
  sequences classification system of][]{:new}.  Incipient central
spiral structure or faint tidal features around a smooth outer disk
may lead to an S0/a classification. The sharp boundary between S0/a
and Sa is of course artificial, particularly in the context of our
interpretation of blue-sequence E/S0s as a transitional
population. Fig~\ref{fg:justone} shows three $g$-band image stretches
for a galaxy we judge to be literally on the borderline between S0/a
and Sa.  We choose to retain such galaxies on the S0/a side of the
division; they make up $\la$10\% of confirmed blue-sequence E/S0s in
the HyperLeda+ sample.

Nothing in the morphological definition of an E/S0 prohibits rings,
bars, gas, dust, star formation, or companion interactions, provided
these do not strongly affect the primary bulge+smooth outer disk
morphology.  Although we largely exclude major mergers in progress by
rejecting HyperLeda+ systems marked as ``multiple,'' our samples
include peculiar E/S0s such as polar ring systems and spheroids with
shells or tails suggesting late-stage mergers.

\begin{figure*}
\epsscale{1.2}
\plotone{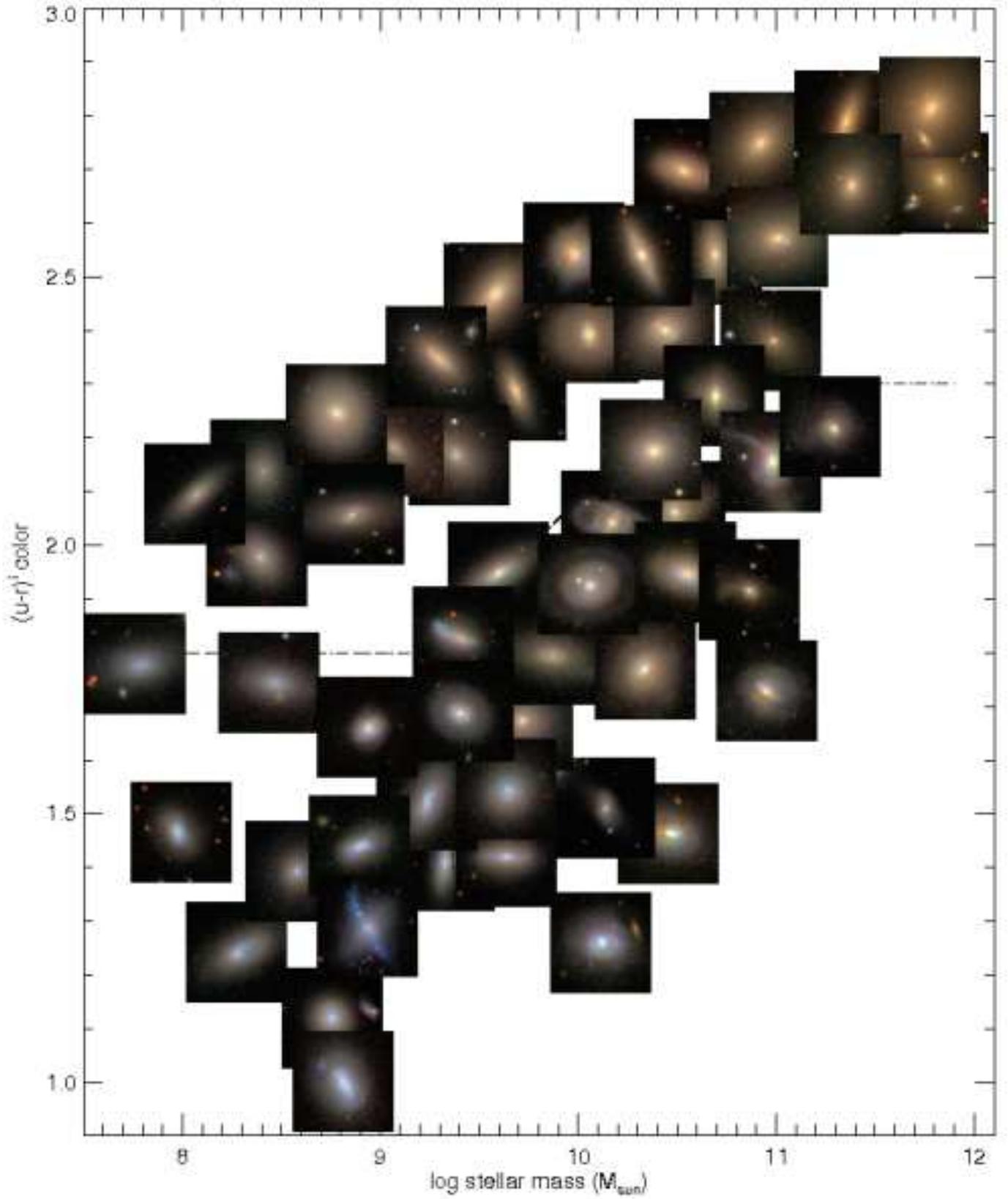}
\caption{SDSS {\it ugriz} color-composite cutouts of E/S0s from both
sequences, arranged by $u-r$ color and stellar mass.  We include
examples from both the HyperLeda+ and NFGS samples, using the $u-r$ to
$U-R$ conversion given in Section~\ref{sec:divider}.  Images are
slightly shifted to minimize overlap and sized arbitrarily.}
\label{fg:montage}
\end{figure*}

Fig.~\ref{fg:montage} shows a representative montage of E/S0 images
from both sequences and both samples.  We stress that the color SDSS
cutout images in Fig.~\ref{fg:montage} do not necessarily reflect the
morphology seen in the monochrome classification images: the cutouts
highlight dust and color structure while downplaying low surface
brightness outer disk features.

Three of the 15 NFGS blue-sequence E/S0s are also in the HyperLeda+
sample, but are not considered E/S0s in the latter sample.  One is
UGC\,9562, the polar ring galaxy shown near ($\log{M_*}$, color) =
(9.2, 1.3) in Fig.~\ref{fg:montage}; because this galaxy is classified
as an Sd in the HyperLeda database, it was never a candidate for
reclassification.  The other two are UGC\,6655 and IC\,1144, which are
listed as E/S0s in the HyperLeda+ database but are among the 60\% of
candidate blue-sequence E/S0s that we rejected upon inspection of SDSS
images.  Reinspection of the NFGS and SDSS images confirms that both
are borderline cases, with the NFGS images making them look more like
S0/a and the SDSS images making them look marginally later in type.
To ensure uniform analysis, we do not change their classifications in
either sample (i.e., they are plotted as late types in the HyperLeda+
sample).  UGC\,6655 is shown in Fig.~\ref{fg:montage} near
($\log{M_*}$, color) = (8.3, 1.5).  These discrepancies highlight the
inexact nature of morphological classification; however,
\S~\ref{sec:scaling} will show that E/S0s in both samples have
structural properties that are quantitatively distinct from later type
galaxies.

\subsection{Stellar Mass Estimation}
\label{sec:massmethods}

Stellar masses are estimated using an improved version of the stellar
population fitting code described in
\citet{kannappan.gawiser:systematic}. The code fits photometry
($UBRJHK$ or $ugrizJHK$) and integrated spectrophotometry (optional:
NFGS only) with a grid of models built up from the simple stellar
population (SSP) models of \citet[][]{bruzual.charlot:stellar} for a
\citet{salpeter:luminosity} Initial Mass Function (IMF).  We rescale
the final masses by a factor of 0.7 to approximate an IMF with fewer
low-mass stars \citep[the ``diet'' Salpeter IMF
of][]{bell.mcintosh.ea:optical}. Each model combines an old SSP (age
1.4, 2.5, 3.5, 4.5... 13.5 Gyr) plus a young SSP (age 25, 100, 290, or
1000 Myr).  The young SSP makes up 0\%, 1\%, 2\%, 4\%, 8\%, 16\%,
32\%, or 64\% by mass; pure young SSPs are also allowed.  Each SSP has
metallicity $Z$ = 0.008, 0.02, or 0.05 and the young SSP can take on
11 extinction/reddening values ($\tau_{V}$ = 0, 0.12, 0.24... 1.2).
Further details are given in \citet{kannappan.gawiser:systematic}.

Following \citet{bundy.ellis.ea:mass*1}, we define the stellar mass
estimate not by the best fit, but by the median and 68\% confidence
interval of the mass likelihood distribution binned over the grid of
models.  The distributions imply typical uncertainties of 0.1--0.2
dex.\footnote{Distance errors are added in quadrature after mass
  estimation, as they are the same for all passbands and therefore
  excluded from the fits.}  In the binned likelihood approach, the
density of models can affect results.  Our current model grid improves
on \citet{kannappan.gawiser:systematic} in that the old population
ages are equally spaced through time, and we weight the model space
such that each of the four young ages contributes 0.25 $\times$ its
model likelihood, with the 11 $\tau_{V}$'s further subdividing the
likelihood contribution.  Note that having included $\tau_V$ in the
models, we do not apply inclination-based internal extinction
corrections to the input photometry used for mass estimation; however,
the output likelihood distributions do not strongly constrain
$\tau_V$, so inclination-based extinction corrections remain the best
option for computing $(U-R)^i$ (\S~\ref{sec:data}).

\begin{figure}
\plotone{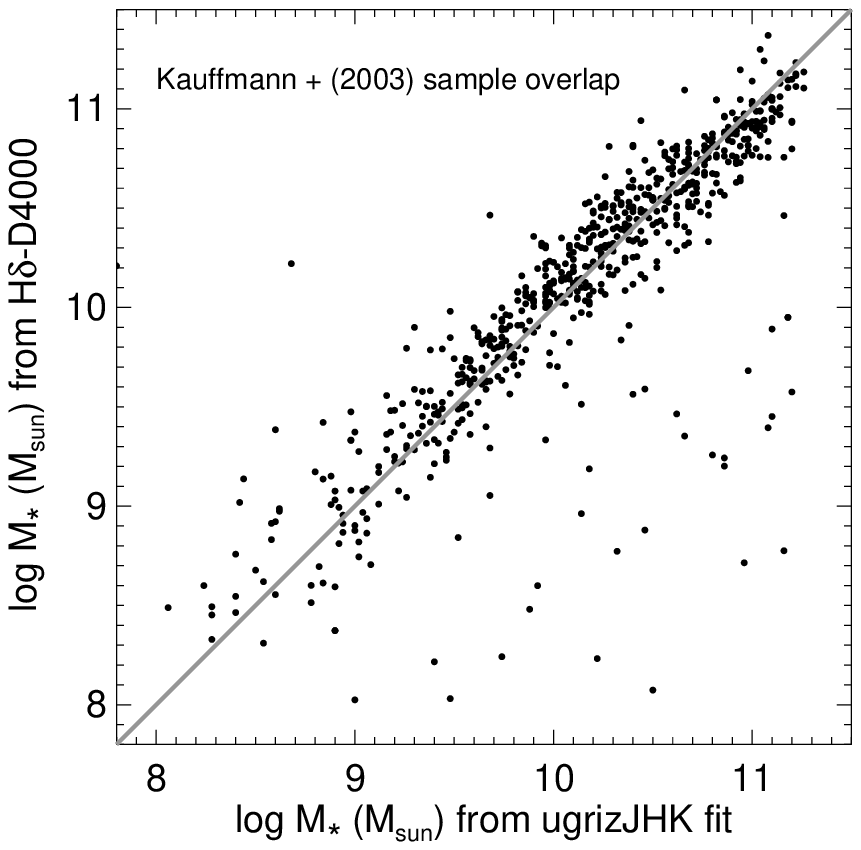}
\caption{Comparison of stellar mass estimates from our $ugrizJHK$
  fitting code and from \citet{kauffmann.heckman.ea:dependence}, for
  658 galaxies in common between the HyperLeda+ and Kauffmann et
  al.\ samples.  The normalization is very similar.}
\label{fg:masscmp}
\end{figure}

\citet{kannappan.gawiser:systematic} report that systematic errors
between different methods of stellar mass estimation can be as large
as factors of 2--3, even with matched IMFs.  Our mass normalization is
fortuitously similar to that of
\citet{kauffmann.heckman.ea:dependence}, as seen in
Fig.~\ref{fg:masscmp}.  Therefore the bimodality mass identified by
Kauffmann et al.\ at 3 $\times$ 10$^{10}$ \msun has roughly the same
value in this paper.  This absolute mass scale is also similar to that
found using the original Kannappan \& Gawiser code with
Bruzual-Charlot models, and as such is intermediate between the lower
and higher mass scales found using either the
\citet{maraston:evolutionary} models or the color-$M/L$ calibration of
\citet{bell.mcintosh.ea:optical}, respectively.  Thus our absolute
scale is a reasonable compromise.  However, the relative position of
red- and blue-sequence E/S0s vs.\ spirals in $M_*$-$r$ and
$M_*$-$\sigma$ scaling relations may be sensitive to systematics in
mass estimation {\it between} galaxy classes, where relative zero
points between classes may vary with estimation method, and the
possible effects of such systematics are discussed in
\S~\ref{sec:scaling}.

\subsection{The Red/Blue Sequence Dividing Line}
\label{sec:divider}

As shown in Fig.~\ref{fg:cmtboth}, we divide the red and blue
sequences with a line that closely hugs the locus of late-type
galaxies, leveling at $(u-r)^i \sim 1.8$ and 2.3 in agreement with the
functional divider of \citet{baldry.glazebrook.ea:quantifying}.  The
NFGS line is shifted to the 0.66 mag bluer $(U-R)^i$ scale and the
0.04 dex lower mass scale that result from using $UBRJHK$+spectra
rather than $ugrizJHK$ as input data, based on empirical calibration
using NFGS galaxies with SDSS photometry.  We measure the 0.66 mag
offset from E/S0s alone to best align the sequence division between
samples; including Sp/Irr types gives a slightly larger 0.75 mag color
offset.

Additional E/S0 systems, possibly analogous to blue-sequence E/S0s,
lie in the minimally populated zone between the sequences (hereafter,
``mid-sequence'').  We separately track the two mid-sequence NFGS
E/S0s seen in Fig.~\ref{fg:cmtboth}b near ($\log{M_*}$, color) = (9.4,
1.3), while for the HyperLeda+ sample, mid-sequence E/S0s are grouped
with red sequence E/S0s.

\section{Demographics and Scaling Relations of Blue-Sequence E/S0s}
\label{sec:exist}

Having robustly identified a population of blue-sequence E/S0s in two
independent surveys, we now examine how this population compares with
red-sequence E/S0s and late types in the overall galaxy population.

\subsection{Numbers, Masses, and Environments}
\label{sec:massenvt}

Raw NFGS numbers give the impression that blue-sequence E/S0s
become dramatically more abundant with decreasing mass
(Fig.~\ref{fg:cmtboth}), especially below $log{M_b/\msun}\sim10.5$.
Here we take a closer look at these galaxies' mass and environment
distributions.

\begin{figure*}
\plotone{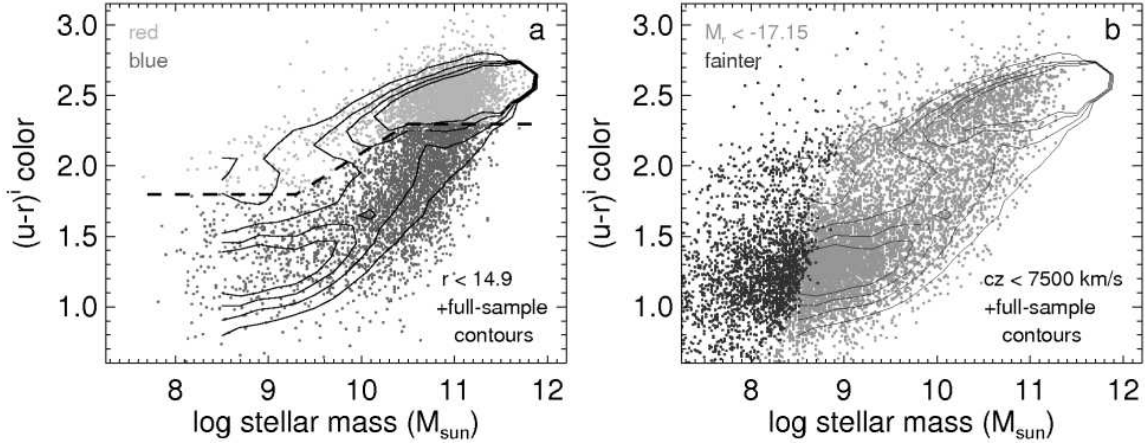}
\caption{Subsamples of the NYU VAGC low-redshift sample with
  well-defined statistical properties. Contours in both panels are
  derived from the full sample (not shown), which spans redshifts
  1000--15000 \kms and has an apparent magnitude limit of $r=18$.
  ({\it a}) Apparent-magnitude limited VAGC subsample with $r<14.9$.
  Light and dark gray points show red- and blue-sequence galaxies,
  respectively.  ({\it b}) Volume-limited sample derived by including
  all VAGC galaxies down to $M_r=-17.15$ with an upper redshift limit
  of 7500 \kms\ (light gray points).  Dark gray points show fainter
  galaxies, demonstrating that this volume-limited sample is largely
  complete in stellar mass down to $\log{M_*/\msun}\sim9$.}
\label{fg:cmpsmp}
\end{figure*}

\begin{figure*}
\plotone{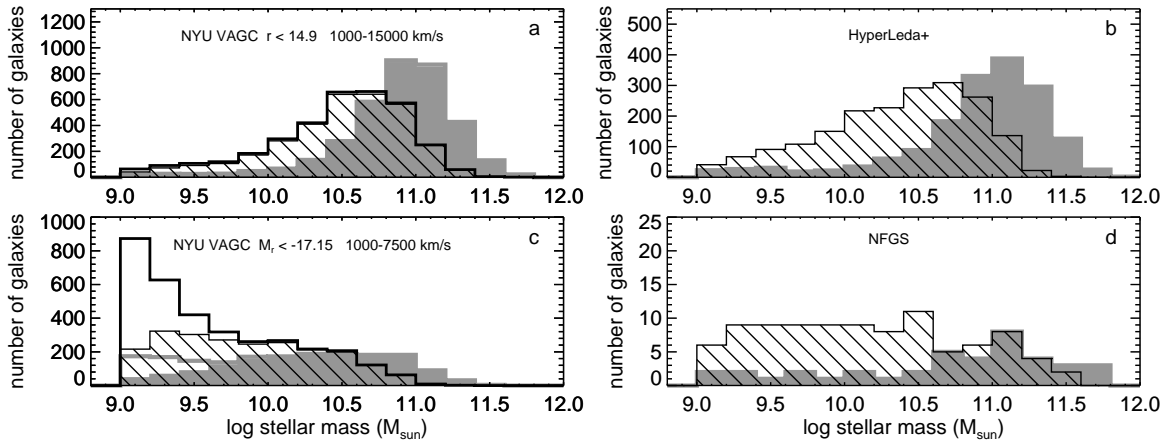}
\caption{Comparison of stellar mass distributions for the two VAGC
  samples of Fig.~\ref{fg:cmpsmp} and the HyperLeda+ and NFGS
  samples. Red- and blue-sequence galaxies are shown in gray and
  black, respectively. For the two VAGC samples, open histograms show
  all selected galaxies, while filled/cross-hatched histograms show
  the subset having a 2MASS $K$-band flux with error $<$0.3 mag,
  matching the selection criteria used for the HyperLeda+ and NFGS
  samples (\S~\ref{sec:samples}).}
\label{fg:cmpsmp2}
\end{figure*}

We compare the HyperLeda+ sample with the NYU VAGC in this analysis,
where both are needed because the VAGC has a simple selection function
without morphological type data, while the HyperLeda+ sample has
morphological type data without a simple selection function.
Fortunately, the HyperLeda+ sample closely resembles an apparent
magnitude limited ($r<14.9$) subsample of the VAGC, within similar
redshift limits (Fig.~\ref{fg:cmpsmp}a and
Figs.~\ref{fg:cmpsmp2}a-b).\footnote{Although the VAGC was designed to
  improve on the raw SDSS, it does not restore large angular-size
  galaxies shredded by the SDSS photometric pipeline and therefore
  omitted from the SDSS redshift survey.  This fact presumably
  explains why Fig.~\ref{fg:cmpsmp2}b contains more high-mass galaxies
  than \ref{fg:cmpsmp2}a.  However, the more dominant blue sequence in
  the HyperLeda+ sample likely reflects the long history of
  $B$-selected redshift surveys feeding into the HyperLeda+ database.}
Therefore it is reasonable to treat the HyperLeda+ sample as an
approximate statistical sample, and we can correct for color and
luminosity biases by binning the HyperLeda+ and VAGC data into mass
bins within each color sequence, then computing a correction factor
for each bin, equaling the ratio between the HyperLeda+ counts and the
corresponding VAGC counts for a {\it volume-}limited VAGC subsample
(Figs.~\ref{fg:cmpsmp}b and~\ref{fg:cmpsmp2}c; defined by
$cz=1000$--7500 \kms and $M_r<-17.15$, where the latter equals the
VAGC's $r = 18$ limit at 7500 \kms). This volume-limited subsample is
largely complete in stellar mass to a limit of $\log{M_*/\msun}=9$, as
seen in Fig.~\ref{fg:cmpsmp}, so we compute number statistics to this
limit.

\begin{figure*}
\plotone{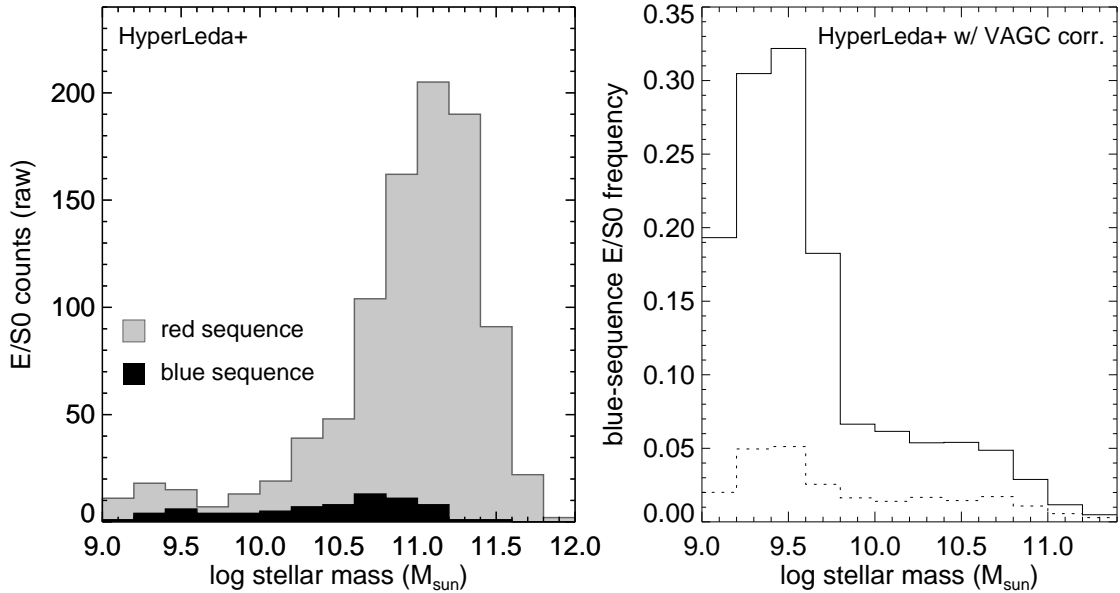}
\caption{Frequency of red- and blue-sequence E/S0s as a function of
  stellar mass in the HyperLeda+ sample. {\it (a)} Raw numbers of
  objects.  {\it (b)} Percentages of blue-sequence E/S0s among all
  E/S0s (solid line) and among all galaxies (dotted line), using the
  NYU VAGC low-redshift catalog to completeness-correct the HyperLeda+
  sample as described in \S~\ref{sec:massenvt}.}
\label{fg:massdist}
\end{figure*}

Fig.~\ref{fg:massdist} presents both the raw HyperLeda+ red- and
blue-sequence E/S0 counts and the completeness-corrected frequency of
blue-sequence E/S0s as a function of stellar mass.  The fraction of
E/S0s on the blue sequence rises steadily with decreasing mass, from
$\sim$2\% at $M_s$ to $\sim$6\% at $M_b$, then shoots up to 20--30\%
below $M_t$.  The overall abundance of blue-sequence E/S0s relative to
the galaxy population rises in tandem, from $<$0.5\% to $\sim$2\% to
$\sim$5\%. We have compared results using completeness corrections
based on the simple volume-limited VAGC sample vs.\ based on a
modified volume-limited subsample restricted by the same $K$-band flux
and error requirements used for the HyperLeda+ and NFGS samples.
Although the $K$-band restrictions reduce overall galaxy numbers below
$\log{M_*/\msun} \sim 9.5$ (Fig.~\ref{fg:cmpsmp2}c), they have minimal
effect on relative numbers between the two sequences.  For
consistency, we present percentages computed with the $K$-band
restrictions.  The decline in blue-sequence E/S0 frequency in the
lowest-mass bin of Fig.~\ref{fg:massdist} should not be
overinterpreted, as shredding and explicit surface brightness
selection criteria start to significantly affect SDSS and thus VAGC
dwarf galaxy counts in this mass bin.

We emphasize that {\it all of these blue-sequence E/S0 frequency
  estimates are lower limits,} both because of our one-sided
morphological reclassification effort (\S~\ref{sec:etgdef}) and
because mid-sequence E/S0s are grouped with the red sequence in the
HyperLeda+ sample (\S~\ref{sec:divider}).  The uncorrected NFGS, which
behaves somewhat like the $K$-band restricted version of the VAGC
volume-limited subsample (Fig.~\ref{fg:cmpsmp2}d), shows higher
blue-sequence E/S0 fractions.  However, these numbers may be affected
by the survey's $B$-band selection, and the small number statistics
and complex selection function of the NFGS make
completeness-correcting it prohibitive.

Interestingly, Fig.~\ref{fg:massdist} hints at a connection between
red- and blue-sequence E/S0s below $M_b$: while the two E/S0 families
have very different mass distributions over all masses (K-S test
probability $2\times10^{-12}$ of being drawn from the same
population), their mass distributions below $M_b$ are
indistinguishable (K-S test probability 0.9 of being drawn from the
same population).  We suggest that evolutionary processes affecting
{\it both} red- and blue-sequence E/S0s may be changing across this
mass scale, plausibly linking the emergence of large numbers of
blue-sequence E/S0s to the onset of high cold gas fractions, bursty
star formation, and disky morphologies, starting around $M_b$ and
becoming more pronounced below $M_t$
\citep{kauffmann.heckman.ea:dependence,kannappan:linking}.

\begin{figure}
\plotone{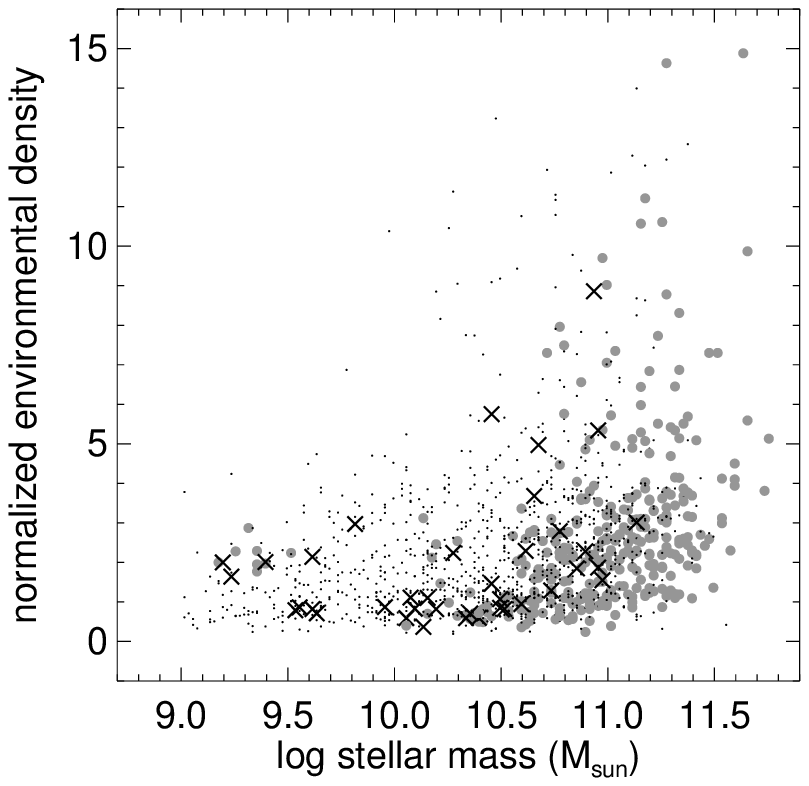}
\caption{Distribution of HyperLeda+ and NFGS galaxies in normalized
  global environmental density vs.\ stellar mass.  Black crosses mark
  blue-sequence E/S0s, gray dots mark red-sequence E/S0s, and small
  black dots mark late-type galaxies.  Densities are expressed in
  units of the mean density of galaxies brighter than $M_B\sim-17$ in
  the Updated Zwicky Catalog \citep[][]{falco.kurtz.ea:updated},
  smoothed on $\sim$7 Mpc scales, using code adapted from N.\ Grogin
  \citep{grogin.geller:lyalpha}.  In these units the densities of the
  Virgo and Coma clusters are $\sim$4.9 and 7.4, respectively.  To
  minimize edge effects, we plot only galaxies lying more than one
  smoothing length from the edge of the UZC and having redshifts
  $cz>1000$ and $<$ 9500 \kms\ with respect to the Local Group.}
\label{fg:envtdist}
\end{figure}

This suggestion agrees with the available environmental data for red-
and blue-sequence E/S0s in the $log{M_*/\msun}=9$--10.5 mass regime,
where most E/S0s seem to occupy similar, and notably low-density,
environments (Fig.~\ref{fg:envtdist}).  Completeness corrections as a
function of environment are infeasible for either of our samples, but
within these data sets, the low-density environments of blue-sequence
E/S0s appear correlated with their intermediate masses.  This result
finds support in the SDSS analysis of
\citet{hogg.blanton.ea:overdensities}, who report a non-monotonic
density trend along the red sequence, seen as a dip in typical density
for intermediate-mass red-sequence galaxies despite higher densities
at higher and lower masses (the latter likely due to the rise of dEs
below 10$^8$--10$^9$ \msun).  Contrary to common assumption, the
environments of intermediate-mass E/S0s are {\it not} necessarily more
dense than those of spiral galaxies, as seen in both
Fig.~\ref{fg:envtdist} and \citet{hogg.blanton.ea:overdensities}.
Neither result is immune to cosmic variance, but both demonstrate that
intermediate-mass E/S0s are abundant in low-density environments where
disk regrowth is plausible.  Analysis of group dynamics would be
beneficial for understanding the balance of cold gas accretion vs.\
quenching processes in blue-sequence E/S0 environments, but such an
analysis is beyond the scope of this paper.  We do consider evidence
that interactions play a role in blue-sequence E/S0 evolution in
\S~\ref{sec:morph} and \S~\ref{sec:interac}.

Taken together, Figs.~\ref{fg:massdist} and~\ref{fg:envtdist} may
explain why blue-sequence E/S0s have been missed as an important
population up to now, despite their large numbers.  First, they emerge
in numbers competitive with red-sequence E/S0s only below $M_t$, so
inevitably they are underrepresented in magnitude-limited samples.
Second, blue-sequence E/S0s are rare in the dense cluster environments
often targeted for studies of early-type galaxies.

\subsection{$M_*$-$r$ and $M_*$-$\sigma$ Scaling Relations}
\label{sec:scaling}

\begin{figure*}
\plotone{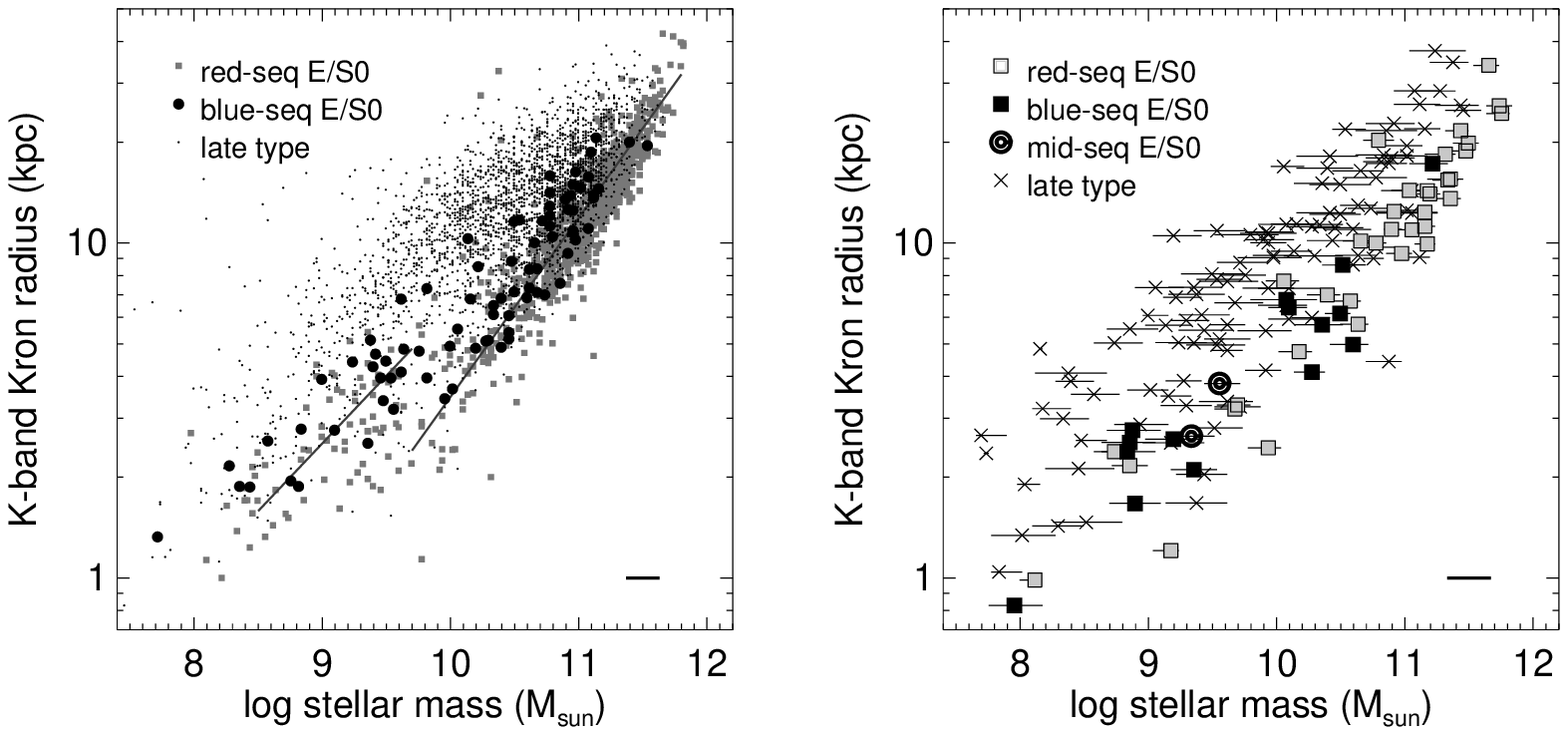}
\caption{$K$-band radius vs.\ stellar mass for {\it (a)} the
  HyperLeda+ sample and {\it (b)} the NFGS sample. The horizontal bar
  represents a typical 68\% confidence interval in mass (i.e., from
  $-$1$\sigma$ to +1$\sigma $). We use 2MASS Kron radii to emphasize
  underlying stellar structure; these radii lack catalogued
  uncertainties but do not seem to contribute much scatter.  Lines in
  panel {\it a} represent least-squares bisector fits for E/S0s above
  and below $M_t$, illustrating the offset of blue-sequence E/S0s
  toward larger radii as well as the general shift in E/S0 locus below
  the threshold mass.}
\label{fg:massradius}
\end{figure*}

\begin{figure*}
\plotone{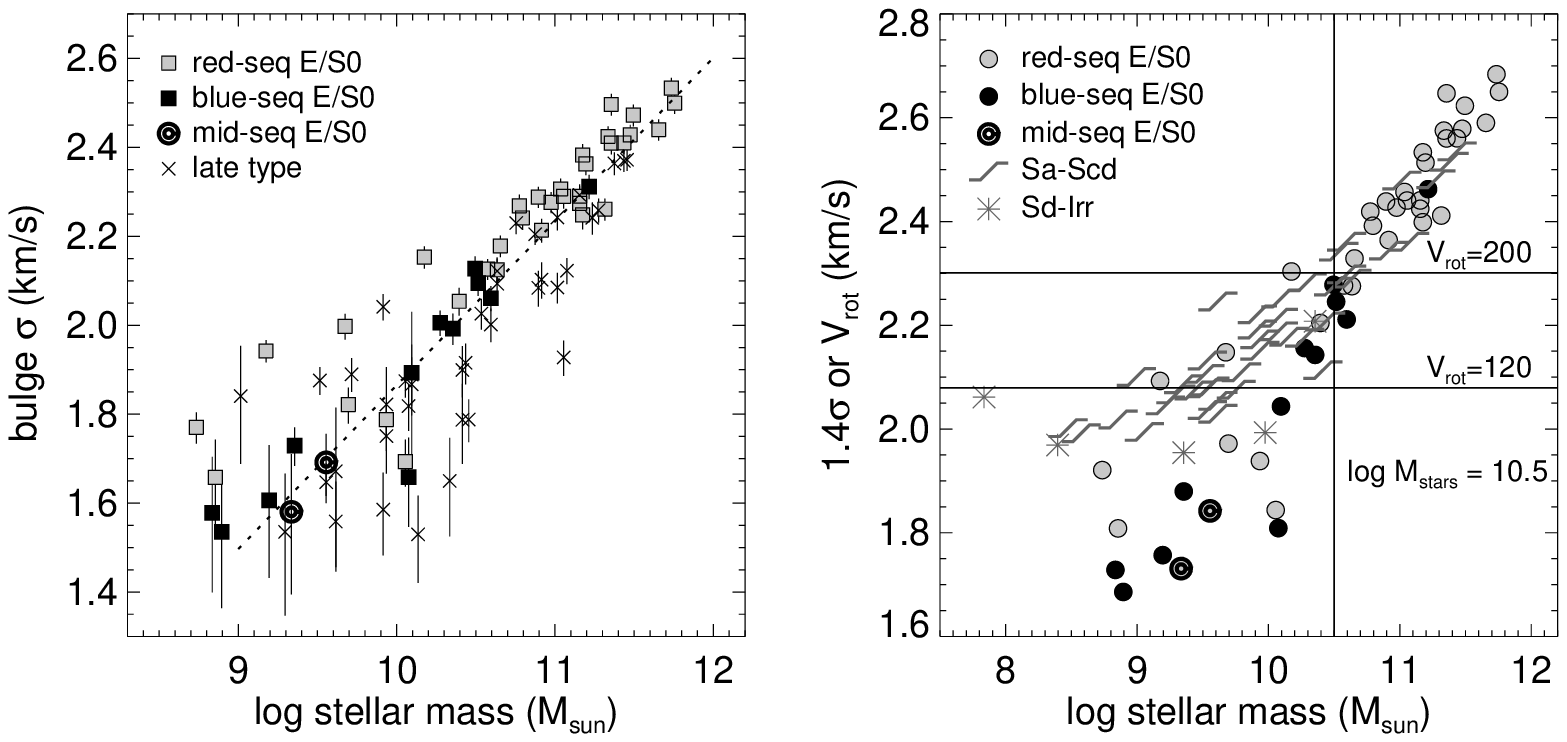}
\caption{Dynamical scaling relations. {\it (a)} Stellar mass
  vs.\ central velocity dispersion $\sigma$ for the NFGS sample.
  Velocity dispersions are measured within an $r_e/4$ aperture.  We
  exclude measurements with formal uncertainties $>$70 \kms\ or
  $>$$\sigma/2$ and add 5\% in quadrature (typical repeatability) to
  the formal uncertainties to compute the error bars shown. The line
  is a least-squares bisector fit to all the data. {\it (b)} Stellar
  mass vs.\ internal velocity for the NFGS sample. E/S0s are plotted
  with velocity dispersions as in panel {\it a}, here multiplied by
  $\sqrt{2}$ to scale with rotation velocities
  \citep[e.g.,][]{burstein.bender.ea:global}.  Late-type galaxies are
  plotted with rotation velocities, equal to half the
  inclination-corrected $W_{V_{pmm}}$ measure of
  \citet[][]{kannappan.fabricant.ea:physical}. Strongly asymmetric or
  truncated rotation curves are excluded by the criteria of
  \citet{kannappan.barton:tools}. Lines mark key mass and velocity
  scales, as indicated.}
\label{fg:masssigma}
\end{figure*}

Blue-sequence E/S0s are more similar to red-sequence E/S0s than to
late-type galaxies in both the $M_*$-radius relation and the
$M_*$-$\sigma$ (stellar velocity dispersion) relation
(Figs.~\ref{fg:massradius} and~\ref{fg:masssigma}).  This basic result
validates their morphological classification.  At the same time, both
relations show hints that blue-sequence E/S0s may represent a
transitional class, and that E/S0s on both sequences change in
structure below $M_b$--$M_t$.  The two NFGS E/S0s from the mid-sequence
zone between the red and blue sequences appear most consistent with
blue-sequence E/S0s in the $M_*$-$r$ and $M_*$-$\sigma$ relations, but
these mid-sequence systems fall within the scatter for both families of
E/S0s.

In the mass-radius relation, blue-sequence E/S0s scatter toward larger
radii than red-sequence E/S0s at fixed mass.  Moreover, many red- and
most blue-sequence E/S0s follow a shifted mass-radius relation below
$\log{M_t/\msun}\sim9.7$ (Fig.~\ref{fg:massradius}), in the same regime
where blue-sequence E/S0s become abundant.  The trend toward larger
radii for blue-sequence E/S0s is not dramatic in this regime --- below
$M_t$, its K-S test significance is only $\sim$2$\sigma$ --- but it is
reproduced in both samples (Fig.~\ref{fg:massradius}).  To minimize
the effect of stellar population differences on this result, we have
adopted the $K$-band elliptical Kron radius from the 2MASS database as
our size measure.  This radius is typically $\sim$2.5 $\times$ the
elliptical half-light radius (and is in fact defined as 2.5 $\times$
the first moment of the light distribution), but in practice
$r_{Kron}^K$ displays less scatter than $r_e^K$ in the mass-radius
relation.

In the mass-$\sigma$ relation, scatter increases dramatically below
$M_b$ (Fig.~\ref{fg:masssigma}), most notably for spirals and
red-sequence E/S0s.  Comparing residuals relative to the overall fit,
K-S tests show that spirals and blue-sequence E/S0s are drawn from
different parent populations at $\ga$3$\sigma$ confidence, whereas
red- and blue-sequence E/S0s are consistent with the same parent
population.  Counting mid-sequence objects as blue- rather than
red-sequence E/S0s slightly differentiates the two types of E/S0,
yielding $\sim$17\% probability that both are drawn from the same
population below $M_b$.  To the extent that there are differences in
this mass regime, blue-sequence E/S0s seem to have low stellar
velocity dispersions more consistently, or equivalently, seem to lack
scatter toward high dispersions.  We suspect that the higher scatter
in red-sequence $\sigma$ values points to greater dynamical diversity,
especially since this scatter is not driven by errors: the four
high-dispersion red-sequence E/S0s below $M_b$ all have very small
errors on $\sigma$.  However, a larger sample would be needed to
confirm the scatter difference statistically, and high-quality
kinematic data sensitive to small $\sigma$ values are not available
for the HyperLeda+ sample.\footnote{SDSS velocity dispersions are
  limited to $\log{\sigma}\ga1.8$--1.9 due to instrumental resolution.
  Within slight differences in details of $M_*$ and $\sigma$
  estimation, our $M_*$-$\sigma$ relation agrees with that of
  \citet{gallazzi.charlot.ea:ages} above $M_b$, but we find higher
  scatter in $\sigma$ at lower masses.}

The mass-$\sigma$ relation of Fig.~\ref{fg:masssigma} is {\it not} the
Faber-Jackson relation between bulge mass and velocity dispersion, but
rather the global relation between {\it total} stellar mass and
central velocity dispersion.  A galaxy's position in this global
relation may depend on bulge-to-disk ratio and on how much the system
is supported by rotation, including whether the bulge is dynamically
cold, as expected for a disky ``pseudobulge'' formed by disk gas
inflow and central star formation \citep{kormendy.kennicutt:secular}.
For comparison, the right panel of Fig.~\ref{fg:masssigma} shows the
global relation between internal velocity and stellar mass, adopting
the simplistic assumption that spirals are rotation-supported and
E/S0s are dispersion-supported, and scaling the dispersions to
equivalent circular velocities by a factor of $\sqrt{2}$
\citep[e.g.,][]{burstein.bender.ea:global}.  The four high-dispersion
red-sequence E/S0s below $M_b$ follow the relation defined by spiral
galaxies and high-mass E/S0s, whereas the lower-dispersion E/S0s fall
short in dynamical support, implying that they are probably supported
by rotation \citep[as is common for E/S0s in this mass
range;][]{davies.efstathiou.ea:kinematic,bender.burstein.ea:dynamically*1}.
Unfortunately, in most cases the rotation velocities of the
low-dispersion E/S0s are impossible to measure with our existing data,
due to abundant warps and asymmetries in their ionized-gas and stellar
rotation curves as well as frequent radial truncation of the ionized
gas emission lines (see also \S~\ref{sec:interac}).

A tendency toward larger radii in blue-sequence E/S0s can be
interpreted as consistent with either evolution from a more compact
state via disk building or incomplete evolution toward greater central
concentration, as in a post-merger starburst phase.  Lower $\sigma$'s
for blue-sequence E/S0s would be more naturally explained in the
disk-building scenario, assuming that in the post-merger scenario the
galaxy should have experienced dynamical heating
\citep[e.g.,][]{dasyra.tacconi.ea:dynamical}, but a gas-rich merger
could involve immediate inner disk regrowth via inflows.

These results rely on stellar mass estimates that may vary
systematically between subpopulations with different characteristic
star formation histories, depending on the methods and assumptions
used \citep{kannappan.gawiser:systematic}.  Substituting stellar
masses derived from the color-$M_*/L$ calibration of
\citet{bell.mcintosh.ea:optical} yields greater scatter in the
$M_*$-$\sigma$ and $M_*$-internal velocity relations and more strongly
differentiates red- and blue-sequence E/S0s in these relations.  The
offset in radii between the two populations diminishes under this
substitution but does not disappear.  Conversely, adoption of the
stellar masses derived by Kannappan \& Gawiser using
\citet{maraston:evolutionary} population synthesis models enhances the
mass-radius offset while diminishing the mass-$\sigma$ offset.
However, revised mass calculations using the more representative model
grid of star formation histories described in \S~\ref{sec:massmethods}
yield very similar results for Maraston and Bruzual-Charlot models.

\section{Characteristics of Blue-Sequence E/S0s}
\label{sec:props}

Here we offer an overview of blue-sequence E/S0 properties, with
special attention to properties relevant to evaluating the
disk-building and fading-merger scenarios.

\subsection{Morphologies}
\label{sec:morph}

Blue-sequence E/S0 morphologies depend strongly on mass
(Fig.~\ref{fg:montage}).  At high masses between $M_b$ and $M_s$
($\log{M_*/\msun}=10.5$--$11.2$), many blue-sequence E/S0s resemble
interacting galaxies or recent merger remnants.  Of the 83
blue-sequence E/S0s in the HyperLeda+ sample, we assigned seven the
type ``pec-e'' to indicate spheroidal but otherwise unclassifiably
disturbed morphology, and all seven have masses between $M_b$ and
$M_s$.  The 16 highest-mass blue-sequence E/S0s include five with
``pec-e'' morphology and two more with prominent tidal features or
dust lanes. The $M_b$--$M_s$ population is best probed by the
HyperLeda+ sample because of its top-heavy luminosity distribution,
but we find similar evidence for tidal features and interactions in
two of the four highest-mass blue-sequence E/S0s in the NFGS
sample. At the same time, this mass range is also where the red and
blue sequences converge, and some high-mass blue-sequence E/S0s in the
HyperLeda+ sample appear quite normal; these may simply be
red-sequence E/S0s affected by scatter in $u-r$.

At lower masses where blue-sequence E/S0s are common, strongly
disturbed morphologies are rare.  Most systems have fairly settled,
disky morphologies.  Faint tidal features and evidence of satellite
interactions seem to be frequent but are hard to quantify (these
features are not generally visible in the small, low-contrast cutouts
in Fig.~\ref{fg:montage}); two quantitative but indirect measures of
external disturbance are discussed in \S~\ref{sec:interac}. In the
NFGS sample, nearly all E/S0s with $M_*$ $<$ $M_b$ have S0--S0/a
types, regardless of sequence (Fig.~\ref{fg:cmtboth}b), as most Es
fall on the high-mass red sequence \citep[consistent with the E and S0
  luminosity functions, e.g.,][]{binggeli.sandage.ea:luminosity}.  In
the HyperLeda+ sample we find no unambiguous Es among 83 confirmed
blue-sequence E/S0s (four galaxies are ambiguous between E and S0).
Vetted types are not available for the red sequence for this sample.

An analysis of the detailed morphologies of E/S0s is beyond the scope
of this paper, but we note two possibly interesting facts that merit
further investigation: (1) the ratio between S0 and S0/a types is not
obviously different between the red and blue sequences, based on the
NFGS sample; and (2) bars, rings, and dust are evident in both red- and
blue-sequence E/S0s (Fig.~\ref{fg:montage}).

\subsection{Colors, Gas, and Star Formation}
\label{sec:cgsf}

\begin{figure*}
\plotone{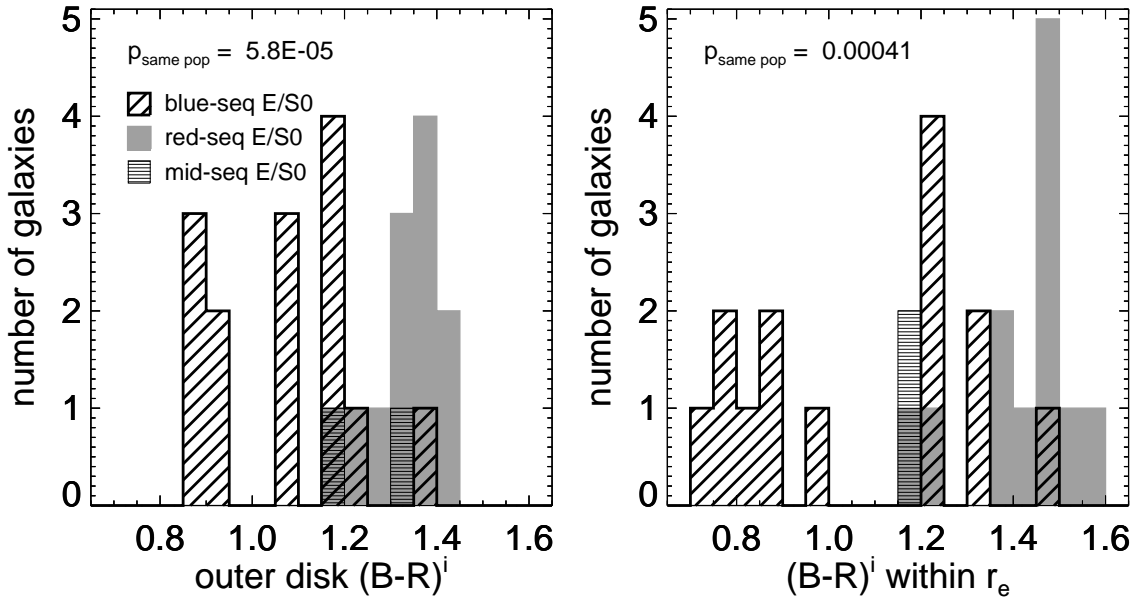}
\caption{Comparison of E/S0 colors for NFGS sample galaxies with
  $\log{M_*/\msun}<10.7$.  Outer disk colors are measured between the
  radii enclosing 50\% and 75\% of the $B$-band light.  Central colors
  are measured within the $B$-band half-light radius.  Each panel
  lists the probability that the red- and blue-sequence samples were
  drawn from the same population according to a Kolmogorov-Smirnov
  test.  We omit NGC\,3605 because of its uncertain $B-R$ color
  profile (\S~\ref{sec:samples}).}
\label{fg:twohists}
\end{figure*}

Intermediate to low mass blue-sequence E/S0s are bluer than
red-sequence E/S0s in {\it both} their outer disks and their centers,
with the strongest statistical difference being for outer-disk colors
(measured between the 50\% and 75\% light radii).
Fig.~\ref{fg:twohists} shows central and outer-disk colors for NFGS
E/S0s, in the mass range below $\log{M_*/\msun}=10.7$ where we have
similar numbers of objects on both sequences.  The two mid-sequence
E/S0s are separated out.  Comparing the central and outer-disk colors
for this subsample, 7 of 14 blue-sequence E/S0s ($50\pm10$\%) have
centers bluer than their outer disks, whereas only 2 of 13
red-sequence E/S0s are blue-centered ($15^{+10}_{-7}$\%).  The
disparity is greater if we limit the comparison to masses below $M_t$,
yielding percentages $86^{+9}_{-16}$\% and $33^{+18}_{-14}$\%,
respectively.  The two mid-sequence E/S0s are both blue-centered, with
outer-disk colors intermediate between those of typical red- and
blue-sequence E/S0s.  Note that these blue central colors are not
caused by AGN: the three known AGN in our sample are all in
red-centered systems \citep[one Seyfert 2 and two LINERs, identified
with NFGS nuclear
spectroscopy:][]{jansen.fabricant.ea:spectrophotometry}.  The high
frequency of blue-centered profiles is intriguing as possible evidence
for interaction-triggered central star formation
(\S~\ref{sec:interac}).  Nonetheless, all of these systems have blue
{\it outer} disks as well, as also seen in blue-sequence E/S0s that
are not blue-centered.

\begin{figure*}
\plotone{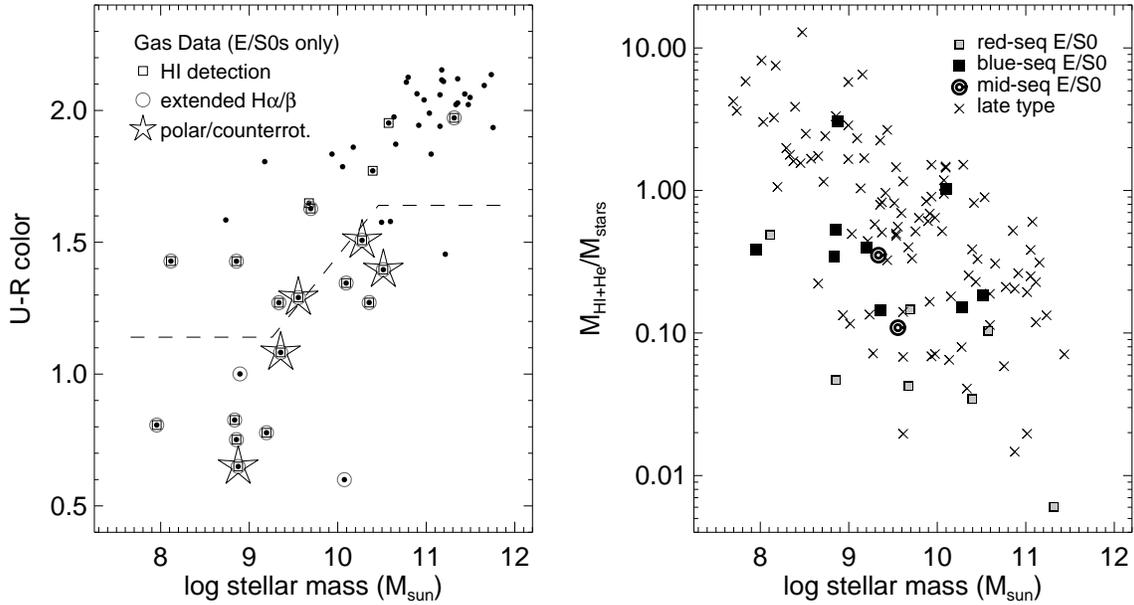}
\caption{{\it (a)} Distribution of gas detections in color-stellar
  mass space for NFGS E/S0s.  Circles indicate optical emission-line
  detections (H$\alpha$ or H$\beta$) in the NFGS database and squares
  indicate 21~cm detections listed in
  \citet{paturel.theureau.ea:hyperleda}.  Galaxies known to have gas
  in counterrotating or polar disks/rings are marked with open
  stars. {\it (b)} Atomic gas-to-stellar mass ratios for NFGS E/S0s
  with HI data.  Late-type galaxies are also shown for comparison.
  Gas fractions are systematically underestimated in the absence of
  molecular gas data.}
\label{fg:gasdata}
\end{figure*}

Fig.~\ref{fg:gasdata} displays gas detections and gas-to-stellar mass
ratios for E/S0s in the NFGS sample.  Stars mark five systems with
counterrotating or polar-ring gas, to be analyzed in detail in
\S~\ref{sec:peckin}.  Although the archival HI data are incomplete and
the ionized gas detections are somewhat biased against red-sequence
E/S0s,\footnote{Spectra for these systems were obtained in a
  wavelength range around H$\beta$ rather than the stronger H$\alpha$
  line.} one can still see a pattern of more frequent gas detections
below $M_t$ on the red sequence and below $M_b$ on the blue sequence,
reflecting an overall increase in gas fractions for the entire galaxy
population \citep[][]{kannappan:linking,kannappan.wei:galaxy}.

At a given stellar mass, blue-sequence E/S0s have more gas than
red-sequence E/S0s, with atomic-gas--to--stellar mass ratios ranging
from $\sim$0.1 to $>$1.  However, the distributions overlap a bit and
do not obviously preclude that some red-sequence and mid-sequence E/S0s
may evolve toward the blue sequence, either gradually or in
intermittent bursts, especially if their environments are favorable to
fresh gas accretion.

\begin{figure*}
\plotone{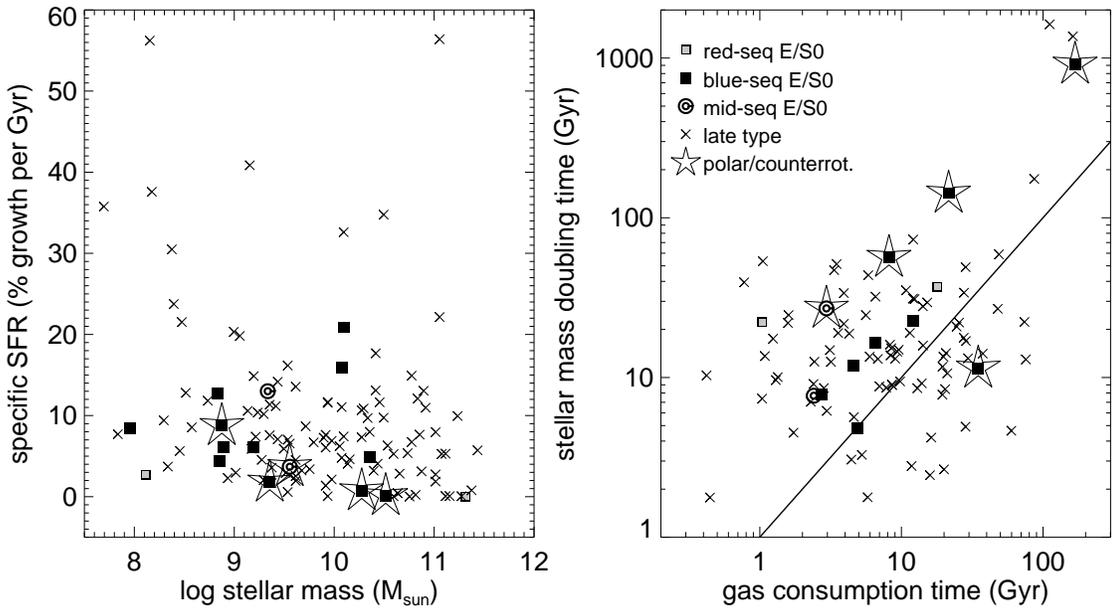}
\caption{Star formation properties of E/S0s. {\it (a)} Specific star
  formation rate as a function of stellar mass for galaxies with
  integrated $H\alpha$ flux measurements in the NFGS database. {\it
    (b)} Stellar mass doubling time vs.\ atomic gas consumption time
  for galaxies with HI data listed in
  \citet{paturel.theureau.ea:hyperleda} and integrated $H\alpha$ flux
  measurements in the NFGS database.  Gas consumption times are
  systematically underestimated due to the neglect of molecular gas
  and fresh infall.}
\label{fg:times}
\end{figure*}

To assess the potential for morphological transformation in
blue-sequence E/S0s, via either disk building or central mass growth,
we consider several measures of star formation summarized in
Fig.~\ref{fg:times}.  We compute the specific star formation rate
(SSFR), expressed in units of percentage growth per Gyr at the current
star formation rate, as well as two star formation timescales: the
stellar mass doubling time, equal to current stellar mass divided by
current SFR, and the atomic gas consumption time, equal to atomic gas
mass divided by current SFR.  These are approximate, instantaneous
measures, uncorrected for any future decline in SFR, recycling, or
infall.  Star formation rates are derived from integrated $H\alpha$
spectral line data from the NFGS database, originally obtained by
scanning the slit across each galaxy
\citep{jansen.fabricant.ea:spectrophotometry}.  We correct for
extinction and convert to the common scale calibrated against
IRAS-based SFRs by \citet{kewley.geller.ea:h}.  Where available, we
adopt the catalogued ``H$\alpha^{corr}$'' SFRs provided by Kewley et
al., and we compute SFRs for additional galaxies in our sample using
integrated H$\alpha$ fluxes provided by R. A. Jansen (private
communication, 2001). All SFRs are converted to our $H_0$ and IMF.

From these data, blue-sequence E/S0s have SSFRs comparable to those of
spiral galaxies (Fig.~\ref{fg:times}a). Rates of transformation range
from $\sim$2--20\% per Gyr.  While SSFRs generally increase toward
lower mass for both spirals and E/S0s, the two most rapidly
transforming E/S0s have masses between $M_t$ and $M_b$. The
highest-mass systems appear quiescent, as do most red-sequence E/S0s,
but one mid-sequence system is evolving at a rate of nearly 15\% per
Gyr.

Most blue-sequence E/S0s, as well as two red-/mid-sequence E/S0s, have
enough gas to form stars for several Gyr.  Fig.~\ref{fg:times}b shows
doubling time vs.\ gas consumption time.  To the left of the line of
equality, galaxies cannot double their stellar masses without fresh
gas accretion (unless the unmeasured molecular gas mass is large).
Four blue-sequence E/S0s and one mid-sequence E/S0 combine doubling
times less than a Hubble time with gas consumption times that are
either longer or $\la$2--3$\times$ shorter, implying significant
evolution even in an unrealistic closed-box scenario.  Notably, the
two-blue sequence E/S0s whose gas consumption times match or exceed
their stellar mass doubling times are both engaged in strong
interactions with smaller, but substantial, companions.

While the above analysis suffers from incomplete data (in particular,
one of the two high SSFR systems in Fig.~\ref{fg:times}a lacks HI data
in in Fig.~\ref{fg:times}b), our results confirm the plausibility of
the basic premise of morphological evolution.  Further analysis of
potential structural changes will require resolved gas and star
formation data, which will soon be available from Spitzer, GALEX, GBT,
CARMA, and VLA programs underway (Wei et al., in prep., and Moffett et
al., in prep.).

\subsection{Interaction Status}
\label{sec:interac}

We have already established the importance of strong interactions for
high-mass blue-sequence E/S0s, based on their morphologies
(\S~\ref{sec:morph}).  However, weaker interactions such as satellite
accretion events are not easy to identify cleanly based on morphology.
Here we offer two other lines of argument to support our subjective
impression that minor interactions are important for blue-sequence
E/S0s below $M_s$.

First, the close correspondence between blue-sequence E/S0s and
blue-centered color gradients found in \S~\ref{sec:cgsf} may reflect
star formation triggered by interactions. In an earlier study of NFGS
galaxies of all types, \citet{kannappan.jansen.ea:forming} showed that
blue-centered color gradients correlate with morphological
peculiarities suggestive of minor mergers and argued that tidal forces
in these accretion events can trigger disk-gas inflows and central
star formation, possibly leading to the formation of disky
``pseudobulges'' \citep{kormendy.kennicutt:secular}.  Most of our
blue-centered E/S0s were not flagged in this prior study as
morphologically peculiar, perhaps because by definition E/S0s are
fairly regular.  Therefore if blue-centered color gradients in E/S0s
reflect recent interactions, the disturbances are likely mild.

As evidence that such mild disturbances may indeed be at work, we find
that below $M_b$, 6 of 11 ($55\pm11$\%) blue-sequence E/S0s with
ionized gas rotation curves in the NFGS database show strong rotation
curve asymmetries \citep[$>$10\% in the asymmetry metric
of][]{kannappan.barton:tools}.  Five of these six are also
blue-centered, although that correspondence is not statistically
significant given the small numbers.  More significantly, the rate of
strong asymmetries for blue-sequence E/S0s is $>$3 times that seen for
the general galaxy population in this mass range ($16\pm2$\%).
Unfortunately, too few red-sequence and mid-sequence E/S0s in our
sample have the extended emission lines necessary to measure rotation
curve asymmetries: only 5 have gas rotation curves, of which 3 are
truncated at $<$0.9$r_e$.

Of course, companion statistics would be valuable in assessing the
role of external disturbances.  While small satellite data are not yet
available, we have identified large companions within 300 kpc and 300
\kms\ of all NFGS galaxies that lie inside the boundaries of the
Updated Zwicky Catalog \citep[UZC;][]{falco.kurtz.ea:updated}.  These
data reveal similar large-companion rates\footnote{Here ``large''
signifies ``as much as $\sim$1 mag fainter,'' because the UZC was
derived from the CfA~2 redshift survey catalog, whose limiting
magnitude is 1 mag fainter than the CfA~1 survey from which the NFGS
was selected.} among red-sequence E/S0s, blue-sequence E/S0s, and
late-type galaxies, subject to data incompleteness.  If anything, the
rate of large companions may be marginally lower among blue-sequence
E/S0s.  Clearly this question would benefit from further investigation.

\section{Counterrotating and Polar Disks:  Secondary Stellar Disk Growth}
\label{sec:peckin}

An especially distinctive feature of blue-sequence E/S0s is their
association with counterrotating gas and polar rings.  As shown in
Fig.~\ref{fg:gasdata}, all five NFGS E/S0s with decoupled gas lie on
or near the blue sequence.  Whether and how these galaxies may form
stars in their decoupled gas is highly relevant to understanding the
evolutionary destiny of blue-sequence E/S0s.  Here we examine these
case studies in some detail, deferring a more general discussion of
formation scenarios to Section~\ref{sec:evol}.

\subsection{The Polar Ring: Triggered Stellar Disk Growth}
\label{sec:pr}

\begin{figure*}
\plotone{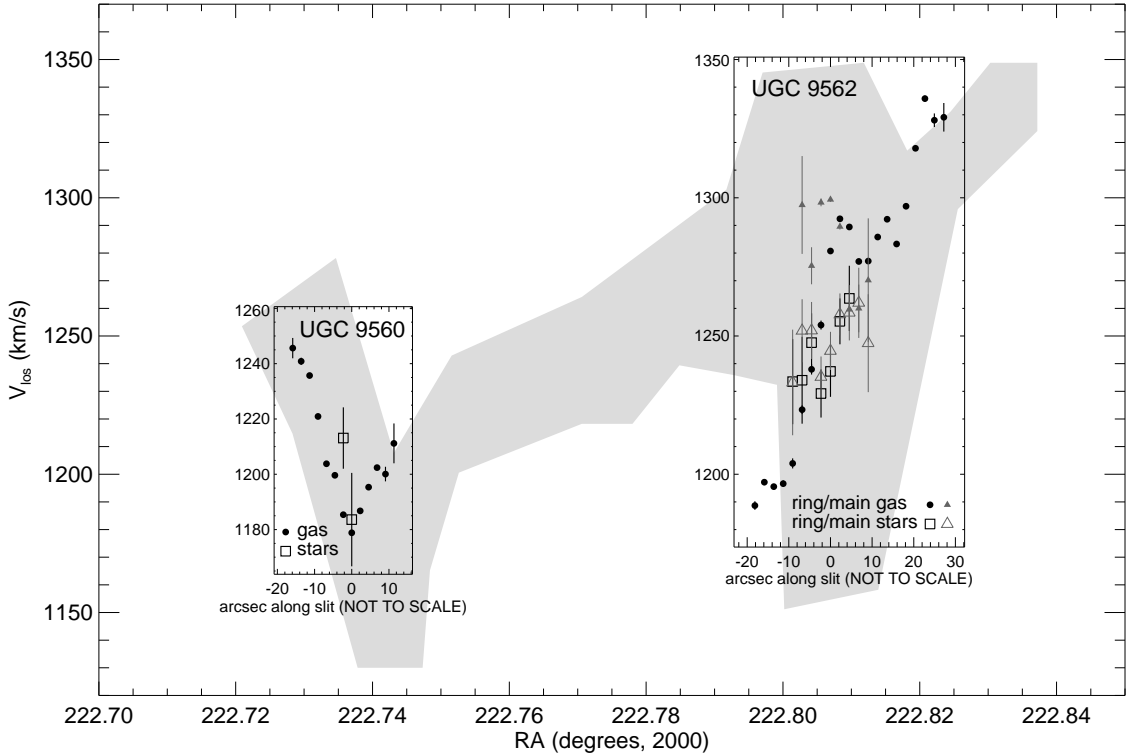}
\caption{Schematic comparison of velocities in different components of
the companion galaxies UGC\,9560 and UGC\,9562 (the polar-ring galaxy
discussed in \S~\ref{sec:pr}).  Grayscale approximates the lowest
surface brightness HI contour detected by \citet{cox.sparke.ea:stars}.
Open triangles and squares mark stellar absorption-line velocities,
while filled triangles and circles mark ionized gas emission-line velocities
from H$\beta$ and the nearby [OIII] lines.  Inset panels are
positioned correctly in position and velocity, but the spatial scales
are expanded to show detail.}
\label{fg:polarring}
\end{figure*}

The polar ring galaxy UGC\,9562 is the bluest and lowest-mass of the
five decoupled gas systems and contains an obvious secondary stellar
disk, i.e., the ring.  This system is shown in Fig.~\ref{fg:montage}
with ($\log{M_*}$, color) near (8.9, 1.3).  UGC\,9562 has recently
interacted with a companion, but \citet{cox.sparke.ea:stars} argue
based on stellar population analysis and HI gas morphology that the
interaction has simply triggered star formation in a {\it preexisting}
gas ring.  Fig.~\ref{fg:polarring} shows new evidence for the
preexisting gas scenario, based on data reported in
\citet{guie.kannappan.ea:extending}.  We plot the ionized gas and
stellar kinematics for UGC\,9562 and its companion overlaid on the HI
position-velocity cut along the major axis of the gas between the
galaxies \citep[see Fig.~3 of][]{cox.sparke.ea:stars}.  As observed by
Cox et al., the HI connecting the galaxies appears to form a distinct
structure from the ring.  Moreover, our new data show that near the
center of UGC\,9562, its {\it ionized} gas has higher line-of-sight
velocity than its stars, while the companion's ionized gas and stars
both have somewhat lower line-of-sight velocities than UGC\,9562's
stars.  These observations are not conclusive but most naturally
suggest a recent flyby event in which previously accreted gas was
tugged and disturbed, rather than an event in which the lower-velocity
system directly donated the high-velocity gas.

These results support the preexisting gas scenario, which implies that
(a) a new stellar disk is forming {\it in situ} in gas that has
accreted onto a previously formed E/S0 and that currently dominates
the total mass ($M_{H,\rm I+He}/M_*\sim 3$), and (b) the mechanisms
and timescales of gas and stellar disk formation may be distinct, with
stellar disk growth being potentially bursty and possibly requiring a
trigger.  Cosmological simulations suggest a picture in which massive
polar rings may form via cold-mode accretion
\citep{maccio.moore.ea:origin}, consistent with observations that
favor a slow gas accretion scenario for another polar ring, NGC\,4650A
\citep[][]{iodice.arnaboldi.ea:stellar}.

\subsection{The Counterrotators:  Past and Present Disk Growth}
\label{sec:cr}

\begin{figure*}
\plotone{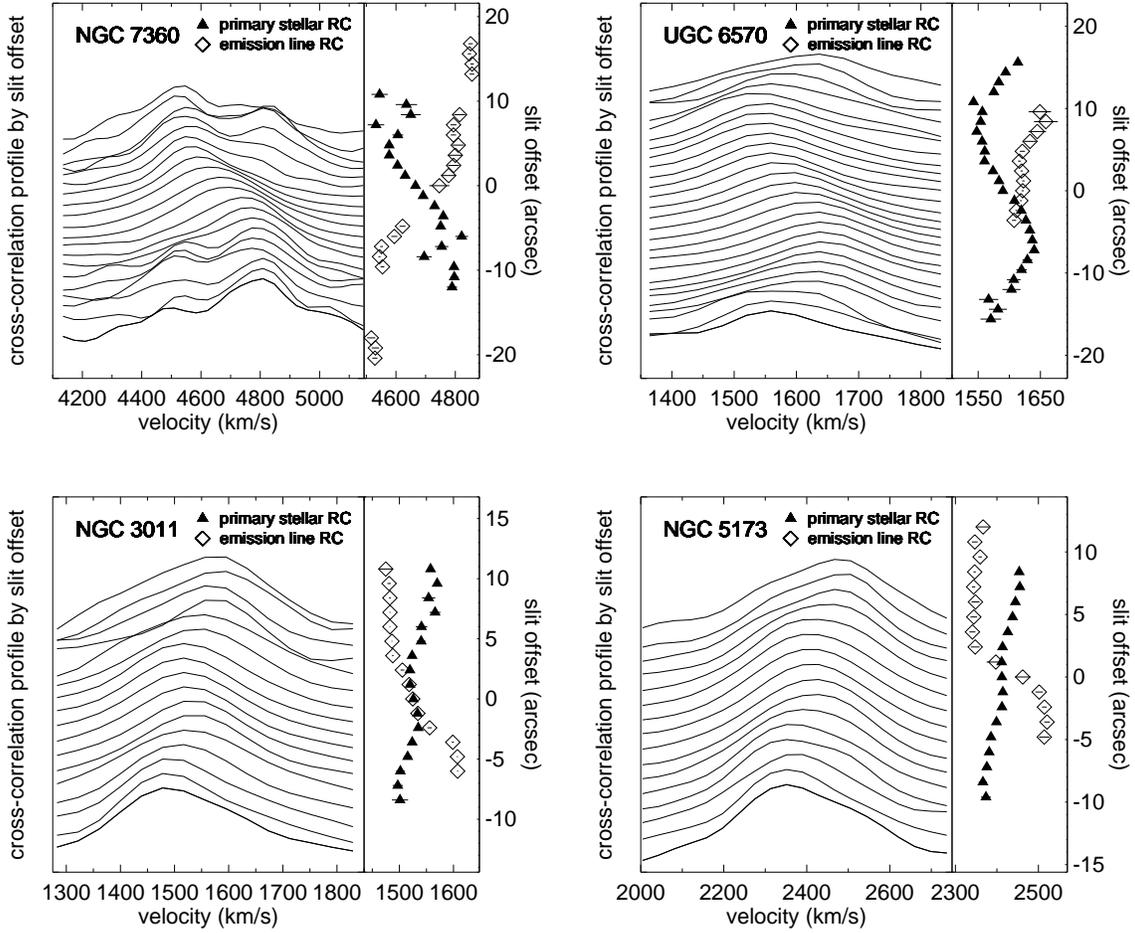}
\caption{Gas and stellar kinematics for four E/S0s identified as
gas-stellar counterrotators by \citet{kannappan.fabricant:broad}.  For
each galaxy, the left panel shows new high S/N stellar absorption-line
cross-correlation profiles at each long slit position, and the right
panel shows the stellar rotation curve extracted from these profiles
(using cross-correlation software that assigns one velocity per
position, \S~\ref{sec:samples}), plus the gas rotation curve extracted
from emission lines in the same long slit spectrum \citep[using
simultaneous emission-line fitting;][]{barton.kannappan.ea:rotation}.
Formal errors are sometimes smaller than the data points.}
\label{fg:rcs}
\end{figure*}

The other four E/S0s marked with stars in Fig.~\ref{fg:gasdata} were
identified by \citet{kannappan.fabricant:broad} in a systematic search
for NFGS galaxies with counterrotating gas and stars.
Fig.~\ref{fg:rcs} shows new, high S/N stellar absorption line
cross-correlation profiles for these galaxies, revealing their stellar
velocity substructure as a function of slit position.  These data were
obtained in May 2001 with the Blue Channel spectrograph on the MMT,
using a configuration with $1\farcs2$ binning and $\sigma\sim40$
km\,s$^{-1}$ resolution centered on the Mg I triplet at 5175\AA. The
spectra are identical to those in the standard NFGS database, except
that deep $\sim$1 hr exposures enable us to search for secondary
stellar velocity components.  We measure the cross-correlation profile
as a function of velocity at each slit position using {\bf xcsao} in
the {\it rvsao} package of IRAF \citep{kurtz.mink.ea:xcsao},
correlating against non-rotating G and K giant stars observed in the
same instrumental configuration. Here we consider what may be learned
from the raw cross-correlation profiles, deferring full modeling of
the line-of-sight velocity distributions.

Secondary stellar components that rotate with the gas are clear in two
systems: (1) the very disky elliptical NGC\,7360, which shows resolved
X-structure in its cross-correlation profiles, indicating oppositely
rotating stellar components; and (2) the low-mass S0 NGC\,3011, which
shows stars and gas that rotate together in the central region, but in
opposite senses at large radii, with profile asymmetries suggesting
that the stellar component that rotates with the gas may be present
but subdominant and unresolved at larger radii.  Profile asymmetries
in a third system, NGC\,5173, likewise suggest the possibility of an
extended, unresolved secondary stellar component rotating with the
gas.  While the asymmetries are not conclusive by themselves, a young
stellar disk is detected morphologically in this galaxy, whose primary
round E component hosts an inclined large-scale spiral disk with blue
knots \citep{vader.vigroux:star-forming}.

In contrast with these three systems, the fourth counterrotator,
UGC\,6570, shows no hint of a secondary stellar component, although its
outer velocities seem to reflect a strong warp or distortion, perhaps
from a recent interaction.  This galaxy may be the exception that
proves the rule regarding secondary stellar disk formation.  We suspect
that its gas arrived too recently to have had time to form stars,
because spectra at different position angles indicate that the gas
rotates in a tilted plane relative to the stars
\citep{kannappan.fabricant:broad}, which is an unstable and
short-lived configuration.  If such an accretion event only recently
pushed UGC\,6570 off the red sequence, then its status as a mid-sequence
E/S0 between sequences may change as star formation proceeds, so that it
may soon join the other polar/counterrotating gas systems on the blue
sequence.

It is empirically clear that three of our four counterrotators have
managed to form significant secondary stellar components, at least one
of which seems to be gaining dominance from the inside out (NGC\,3011),
and another one of which has unmistakably substantial mass (NGC\,7360).
Yet these systems have only $\sim$10--20\% gas fractions and among the
lowest SSFRs in Fig.~\ref{fg:times}.  Bulk accretion of both gas and
stars from an outside galaxy into a disk-like secondary component
seems unlikely to explain these galaxies, considering that the
mid-sequence system UGC\,6570 appears to contain externally accreted
gas without accompanying stars (as we also infer for the polar ring
system in \S~\ref{sec:pr}).  Also, in most of these systems the
counterrotating gas is extended, sometimes well beyond the stars
(NGC\,7360 shows HI and H$\alpha$ emission out to $\ga$3$\times$
its optical radius; Kannappan, Matthews, Christlein, et al., in
prep).  We suspect that the disparity between past and current star
formation reflects a history of intermittent bursts of enhanced
stellar disk growth, occurring independent of gas disk growth.  Also,
the presence of coextensive counterrotating stellar populations may
exacerbate such intermittency, if the galaxy is stabilized against
forming spiral arms or other structures, so that only strong
interactions can revive star formation (Kannappan, Matthews,
Christlein, et al., in prep).

Our detection rate of 2 definite + 1 probable counterrotating stellar
disks in a sample of 4 gas-stellar counterrotators may seem to
contradict the null result of \citet{kuijken.fisher.ea:search}, who
surveyed 28 S0 galaxies for counterrotating stellar disks and found
none.  However, we note that they analyzed only 4 systems with
gas-stellar counterrotation, so the samples that are relevant to
compare are tiny and may differ in crucial parameters such as mass and
color.  The fact that Kuijken et al.\ could not find any
counterrotating stellar disks in S0s {\it without} gas might also have
another explanation: if counterrotating stars reflect disk regrowth,
then S0s with counterrotating stars may be on their way to becoming
spiral galaxies.  If this conversion is efficient, the condition of
still having S0 morphology but having built up a detectable
counterrotating stellar disk may be an unusual state, becoming
long-lived only when the disk regrowth process is quenched
prematurely.

\section{Discussion}
\label{sec:evol}

While individual blue-sequence E/S0s have been studied for years
without being identified as a class, recognizing these galaxies as a
coherent population opens the way to understanding their role in the
story of galaxy evolution.  Below, we break down blue-sequence E/S0
formation scenarios as a function of mass, then take a closer look at
the disk regrowth scenario in light of other studies.  We close by
considering how these results may affect our understanding of galaxy
evolution from $z=1-0$.

\subsection{Evolution in Three Mass Ranges}
\label{sec:threemass}

Taken together, our results suggest that blue-sequence E/S0s are a
composite population, reflecting different physical processes in
different mass regimes.  We distinguish three regimes: above the
shutdown mass $M_s$, below the threshold mass $M_t$, and in the
intermediate range centered on $M_b$. The intermediate regime is the
primary mass regime of large spiral galaxies and as such defines a
coherent population; however, blue-sequence E/S0s in the upper and lower
halves of this regime show distinct evolutionary trajectories.

Above $M_s$, blue-sequence E/S0s are virtually nonexistent, as the
blue sequence itself largely disappears.  From $M_b$ up to $M_s$
($\log{M_*/\msun}\sim10.5$--11.2), there is a clear decline in the
frequency of blue-sequence E/S0s (Fig.~\ref{fg:massdist}), and their
morphologies often show signs of strong disturbance.  Their $\la$1\%
abundance relative to other massive galaxies is roughly comparable to
the frequency of close pairs, suggesting an approximate correspondence
to the merger rate, depending on relative time spent in the close-pair
and fading-starburst phases
\citep[e.g.,][]{de-propris.conselice.ea:millennium,domingue.xu.ea:2masssdss}.
These results point to major and substantial minor mergers as the
probable source of most high-mass blue-sequence E/S0s.  Because cold
gas is generally scarce in E/S0s with masses above $M_b$ --- both
within the galaxies and in their typically dense environments
(Figs.~\ref{fg:envtdist} and~\ref{fg:gasdata}) --- high-mass
blue-sequence E/S0s are likely to exhaust all available gas in their
current post-merger phase and migrate permanently to the red sequence
(``quenching mergers'').  However, limited disk building is also
occurring in this mass regime (e.g., NGC\,7360, \S~\ref{sec:peckin}).

At the other extreme, below the threshold mass
$\log{M_t/\msun}\sim9.7$, blue-sequence E/S0s show little resemblance
to canonical major-merger remnants.  Here blue-sequence E/S0s make up
20--30\% of all E/S0s and $\sim$5\% of all galaxies.  The coincidence
of their emergence with a general rise in gas richness for the entire
galaxy population \citep{kannappan:linking,kannappan.wei:galaxy}
strongly suggests that the availability of cold gas is fundamental to
their existence.

Intermediate-mass E/S0s from $M_b$ down to $M_t$ are often gas rich as
well, at least on the blue sequence, and our two most rapidly growing
E/S0s lie in this mass range (Fig.~\ref{fg:times}).  Available
environmental data reveal a strong shift toward lower density
environments for E/S0s below $M_b$, down to at least 10$^9$ \msun, so
fresh gas accretion seems plausible (Fig.~\ref{fg:envtdist}; recall
that cluster dEs become abundant at lower masses).  The shift in
environments applies to both red- and blue-sequence E/S0s, and the two
populations roughly track each other in numbers below $M_b$
(Fig.~\ref{fg:massdist} and \S~\ref{sec:massenvt}), but red-sequence
E/S0s from $M_b$ down to $M_t$ can be gas poor nonetheless.  In this
mass range, red-sequence E/S0s seem to show greater diversity in
$V/\sigma$ than blue-sequence E/S0s (within small number statistics;
see Figs.~\ref{fg:masssigma}). A shifted $M_*$-radius relation becomes
dominant below $M_t$ for E/S0s on both sequences, but some
red-sequence E/S0s fall on an extension of the higher-mass
$M_*$-radius relation (Figs.~\ref{fg:massradius}).  Blue-sequence
E/S0s make up 5--7\% of all E/S0s and nearly 2\% of all galaxies
between $M_b$ and $M_t$, potentially reflecting evolution in both
directions between the sequences rather than uni-directional
quenching.

\subsection{Forming and Re-Forming Disks}
\label{sec:disks}

Our data suggest that blue-sequence E/S0s in mass and environment
regimes with abundant cold gas are most likely engaged in building
disks.  Blue outer disks are the strongest statistical signature of
blue-sequence E/S0s (\S~\ref{sec:cgsf}), and a number of them show
evidence for secondary stellar disk growth in counterrotating or polar
gas (\S~\ref{sec:peckin}).  At the same time, the fact that
blue-sequence E/S0s are more compact and dynamically hotter than
spirals, and more similar to red-sequence E/S0s than spirals in
scaling relations, suggests that most of these E/S0s have passed
through a violent merger phase at some point in their formation
history (\S~\ref{sec:scaling}).  Other processes that could yield
compact structure, e.g., harassment, are less prevalent in the
modest-density environments seen for the sub-$M_b$ E/S0s in our
samples.  Moreover, these processes would not be expected to produce
blue outer disks or secondary disk growth.  Thus the extended blue
disks seen in blue-sequence E/S0s plausibly represent true disk {\it
  regrowth} over merger-formed early type galaxies, as envisioned in
hierarchical models
\citep[e.g.,][]{steinmetz.navarro:hierarchical,governato.willman.ea:forming}.

A variety of hints in the literature point to the likelihood of disk
regrowth around E/S0s, especially S0 galaxies.  Many S0s contain gas
that appears to have an external origin: HI rings tilted out of the
stellar disk \citep{.:distribution}, polar rings
\citep[][]{bettoni.galletta.ea:gas}, or counterrotating gas
\citep{bertola.buson.ea:external}; see also \citet{sage.welch:cool*1}.
These extended HI disks and rings are often associated with star
formation and/or faint extended stellar disks
\citep[][]{hawarden.longmore.ea:neutral,.schechter.ea:h,noordermeer..ea:westerbork,jeong.bureau.ea:star}.
``Antitruncated'' disks with a low surface brightness component that
dominates at large radii are also reported in S0s
\citep[][]{erwin.beckman.ea:antitruncation}.  Recent star formation in
E/S0s may be substantial, involving $>$10\% of the stellar mass,
especially in field S0 galaxies (\citealt{annibali.bressan.ea:nearby,
  nolan.raychaudhury.ea:young,kaviraj.schawinski.ea:uv-optical,schawinski.kaviraj.ea:effect};
see also \citealt{gallazzi.charlot.ea:ages,donas.deharveng.ea:galex}).
In polar ring galaxies and other systems where secondary disks can be
isolated, these components can be very massive
\citep[\S~\ref{sec:peckin};][]{bettoni.galletta.ea:gas}.

We speculate that disk-building E/S0s may (re)join the spiral sequence
when secondary disks become dominant, perhaps helping to explain the
surprising scarcity of S0s with two counterrotating stellar disks
\citep[\S~\ref{sec:cr};][]{kuijken.fisher.ea:search} and the existence
of peculiar objects like the prolate-bulge Sbc galaxy UGC\,10043
\citep[][]{matthews.:optical}, which could be a former polar
ring galaxy whose ring has become a large disk around the central E/S0.
To grow into spiral galaxies, blue-sequence E/S0s must grow on average
by a factor of $\sim$2 in radius (Fig.~\ref{fg:massradius}),
representing a huge increase in mass.  Thus (re)forming a late-type
system on a cosmologically relevant timescale will require two things:
{\em abundant gas} and {\em efficient star formation}.

Efficient star formation does {\em not} seem to characterize the most
massive gas-rich S0s \citep{oosterloo.morganti.ea:extended}, which
include galaxies such as Malin~1 \citep[recently exposed as an S0 at
  heart by][]{barth:normal} and NGC\,3108 \citep{hau.bower.ea:is}.
Inflow mechanisms appear to be ineffective for E/S0s with dispersions
$\sigma\sim230$ \kms\ \citep{serra.trager.ea:stellar}, equivalent to
$M_*\ga M_s$ (Fig.~\ref{fg:masssigma}).  Nonetheless, minor mergers
may help to drive disk gas inward in intermediate-mass E/S0s, fueling
residual star formation
\citep{kauffmann.heckman.ea:ongoing, kaviraj.peirani.ea:role}.

The real action, however, starts below $M_b$ and increases below
$M_t$, down to $\la$10$^9$ \msun.  Here blue-sequence E/S0s are likely
to have formed in mergers of very gas-rich progenitors (i.e., galaxies
typical of this mass regime: Fig.~\ref{fg:gasdata};
\citealt{kannappan:linking,kannappan.wei:galaxy}).  The remnants would
have inherited substantial gas through the fallback of tidal debris
\citep{barnes:formation,robertson.bullock.ea:merger-driven,stewart.bullock.ea:gas-rich}.
Moreover, the sub-$M_t$ mass regime may permit continued growth by
efficient cold-mode gas accretion
\citep{birnboim.dekel:virial,kere-s.katz.ea:how,dekel.birnboim:galaxy}.
Subsequent interactions can also supply fuel for star formation, via
direct accretion of gas-rich satellites, gas transfers, or compression
of ambient gas.

Star formation becomes remarkably efficient for E/S0s below $M_b$, and
the two most strongly evolving blue-sequence E/S0s in the NFGS, whose
current SFRs correspond to $\sim$20\% mass growth per Gyr, have
$M_*\sim10^{10}\msun$ (Fig.~\ref{fg:times}). This result may seem
counterintuitive, since below $M_t$, most disk galaxies are late-type
dwarfs (Sd, Sm, Im), which have notoriously inefficient star formation
compared to classical large spirals.  However, recent work suggests
that this inefficiency may be linked to low surface mass densities
rather than to low masses {\it per se}.
\citet{kauffmann.heckman.ea:gas} report that evidence of
higher-intensity, shorter starbursts increases with stellar surface
mass density $\mu_*$ for disk-dominated galaxies, reaching a peak at a
characteristic $\mu_*\sim3\times10^8$ \msun\, kpc$^{-2}$ before
falling off for higher-mass/more spheroid-dominated systems as the
star formation rate per unit stellar mass $<$SFR$>$$/M_*$ decreases.
The characteristic $\mu_*$ corresponds to a characteristic
concentration index $C_r\sim2.5$--$2.6$ (where $C_r$ is defined by the
ratio of the radii containing 90\% and 50\% of the light in the SDSS
$r$ band).  Galaxies near the characteristic $\mu_*$ and $C_r$ are
common from $M_t$ up to just below $M_s$ (see Kauffmann et al.'s
Figures 1 and 2).

Intriguingly, blue-sequence E/S0s have nearly optimal $C_r$: based on
SDSS data for 73(50) NFGS galaxies below $M_b$($M_t$), blue-sequence
E/S0s have mean $C_r=2.74$(2.54), with standard deviation $\sim$0.45,
whereas late-type galaxies have $C_r=2.31$(2.28) with standard
deviation $\sim$0.40, and red-sequence E/S0s have $C_r=3.01$(2.86)
with standard deviation $\sim$0.30.  Kauffmann et al.\ interpret the
peak in burst intensities near the characteristic $\mu_*$ in terms of
short gas consumption times for accreted clouds (equivalent to high
star formation efficiencies), because they see fairly constant
$<$SFR$>$$/M_*$ over the range of $\mu_*$ below the characteristic
value.  However, constant $<$SFR$>$$/M_*$ implies relatively lower
$<$SFR$>$$/M_{tot}$ for gas-rich late-type galaxies, a non-trivial
distinction in the sub-$M_t$ regime
\citep[Fig.~\ref{fg:gasdata};][]{kannappan:linking,kannappan.wei:galaxy}.
Therefore, the Kauffmann et al.\ results are consistent with both
higher $<$SFR$>$$/M_{tot}$ and higher-intensity bursts for
blue-sequence E/S0s compared to spiral or red-sequence E/S0 galaxies,
apparently as a direct corollary of having surface mass densities near
the characteristic value.\footnote{Exactly what sets the optimal
  surface density is unclear in the absence of information on gas
  content, $\mu_{gas}$, or the timing of gas accretion relative to
  starbursts. The population-averaged trends seen by Kauffmann et
  al.\ must include a variety of physics, as emphasized by the falloff
  in burst intensities above the characteristic $\mu_*$.  Both local
  disk dynamics and global effects of gas-rich minor or major mergers
  may link higher star formation efficiencies with higher $\mu_*$,
  while quenching mergers may work the other way.}

It would seem that most blue-sequence E/S0s (and likely some
red-sequence E/S0s and non-E/S0 blue compact galaxies) lie in a sweet
spot for disk building, with surface mass densities ideal for
efficient star formation and masses low enough for abundant gas
accretion.  We conjecture that passing through a blue-sequence E/S0 or
similar phase in this sweet spot may enable a galaxy to escape the low
surface density dwarf regime and form a large spiral disk.  This
conjecture is consistent with observations that massive disk galaxies,
no matter how thin or low surface brightness, always seem to show a
bulge component
\citep{dalcanton.yoachim.ea:formation,sprayberry.impey.ea:properties}.
The blue-sequence E/S0s we have analyzed are a natural progenitor
population for large spiral galaxies with masses up to $\sim$$M_b$,
and this evolutionary link is likely to extend to higher masses at
higher $z$ (see \S~\ref{sec:hiz}).  However, the physics of this
transformation, and particularly the growth of the disk beyond the
radius of the original E/S0, is not yet clear.

As the bulges of many late-type galaxies are disky, any late-type
regeneration process should be able to form disky bulges.  We may see
this happening in blue-sequence E/S0s, whose frequent blue-centered
color gradients and kinematic disturbances plausibly reflect episodes
of pseudobulge growth triggered by weak interactions
\citep[\S~\ref{sec:interac}; see also][]{kannappan.jansen.ea:forming}.
Externally or internally driven gas inflows can gradually build a
larger pseudobulge over a classical bulge, as is seen in the detailed
HST decomposition of two S0s \citep{erwin.beltran.ea:when}.  In fact,
pseudobulges are quite common in S0s
\citep[e.g.,][]{laurikainen.salo.ea:morphology}, and
\citet{barway.kembhavi.ea:lenticular} find evidence for a sharp
transition in the scaling between S0 bulge and disk radii below
$M_K\sim-24.5$ (in our system, equivalent to
$\log{M_*/\msun}\sim10.8$--10.9), which they interpret as evidence of
pseudobulge growth.  We suspect that nearly all blue-sequence S0s have
pseudobulges, given their modest velocity dispersions, whereas the
large scatter in red-sequence E/S0 velocity dispersions probably
indicates a variety of bulge types.  Only systems in which the central
spheroid is protected from gas inflows, as in a polar ring
configuration, might preserve the original spheroid while building
just the outer disk (as for the aforementioned prolate-bulge galaxy
UGC\,10043).

The timescale for disk regrowth is probably longer than a typical
merger timescale, given that most blue-sequence E/S0s show relatively
regular morphology, and that their frequency in the galaxy population
exceeds that of close pairs (except near $M_s$; Fig.~\ref{fg:massdist}
and \S~\ref{sec:threemass}).  However, small bursts of growth must be
common, as $\ga$50\% of blue-sequence E/S0s show blue-centered color
gradients and/or kinematic asymmetries (\S~\ref{sec:interac},
\ref{sec:cgsf}, and~\ref{sec:peckin}).  Bursty disk growth appears to
be the norm for large disk galaxies, if the star formation and
accretion histories of the Milky Way and M31 are representative
\citep[][]{rocha-pinto.scalo.ea:intermittent,.:on*1,helmi.navarro.ea:pieces,ibata.chapman.ea:on,elmegreen.scalo:effect}.
In fact, \citet{hammer.puech.ea:milky} argue that the Milky Way's
history may be considered unusually quiet \citep[but
  see][]{james.oneill.ea:h}.  A variety of evidence suggests that disk
galaxies receive regular fresh gas infall sufficient to fuel episodic
bursts \citep{sancisi.fraternali.ea:cold}.

The likelihood of intermittent growth and/or dust-enshrouded star
formation implies that some former or future blue-sequence E/S0s
should fall on the red sequence.  Consistent with this picture, we
find regimes of overlap in red-and blue-sequence E/S0 properties (gas
fractions, radii, velocity dispersions), with substantial diversity
among red-sequence E/S0s.  With existing data, we cannot measure
rotation velocities for the dynamically cold red- and blue-sequence
E/S0s in our sample (\S~\ref{sec:scaling}), but many have stellar
masses sufficient to join the spiral-galaxy Tully-Fisher relation
above $M_t$ (i.e., above $V_{rot}$ $\sim$ 120 \kms), provided they
grow large rotating disks (Fig.~\ref{fg:masssigma}). One such
blue-sequence S0, UGC\,12265N, has enough gas to double its stellar
mass and is engaged in a strong interaction with a smaller but
substantial companion, driving stellar mass growth at a rate of
$\sim$20\% per Gyr (\S~\ref{sec:cgsf}). Because the eventual merger
should be too gas rich to destroy the disk \citep{hopkins.cox.ea:how},
this galaxy will almost certainly develop spiral structure as the disk
forms stars and responds to tidal forces in the interaction.  Such an
event may mark the birth of a large spiral galaxy.

\subsection{Implications for Galaxy Evolution from $z=1-0$}
\label{sec:hiz}

Most of the above discussion refers to a snapshot at $z=0$. Given the
downsizing of the bimodality mass, e.g., as seen in the downsizing of
the crossover mass for E/S0 and late-type mass functions from close to
$M_s$ at $z=1$ to $\sim$$M_b$ today \citep{bundy.ellis.ea:mass*1}, we
suspect that the gas-richness threshold mass $M_t$ has also evolved
from $z=1-0$.  We therefore expect $z\sim1$ analogues of blue-sequence
E/S0s to show strong disk-building activity up to the equivalent
threshold mass at $z\sim1$, which might naively be expected to fall
$\la$1 dex below the $z\sim1$ crossover mass, i.e. near a few $\times$
10$^{10}$ \msun\ (similar to today's $M_b$).  Conversely, a noticeable
fraction of $z=0$ galaxies with masses up to $M_s$ may represent the
end states of high-redshift blue-sequence E/S0s \citep[or dusty
  red-sequence analogues;][]{hammer.flores.ea:did} that have rebuilt
disks.  Consistent with this expectation,
\citet{sargent.carollo.ea:evolution} report from the COSMOS survey
that the number density of large disks with bulges has doubled since
$z=1$, whereas they see no increase, and possibly a small decrease, in
the number density of bulgeless disks at $z=0$.  These results point
to the importance of forming bulges before building large disks, i.e.,
merging gas-rich bulgeless late types to form E/S0s may be an enabling
step in large disk galaxy formation (as also suggested by the star
formation efficiency considerations in \S~\ref{sec:disks}).

Analogues of today's sub-$M_t$ blue-sequence E/S0s, with masses up to
a higher threshold mass perhaps near a few $\times$ 10$^{10}$ \msun,
must certainly exist in high-$z$ samples of blue spheroids, blue-cored
spheroids, and luminous blue compact galaxies
\citep[e.g.,][]{im.faber.ea:are,menanteau.abraham.ea:evidence,treu.ellis.ea:assembly,elmegreen.elmegreen.ea:central,lee.lee.ea:nature}.
Type-dependent luminosity function analyses reveal abundant blue
E/S0s, whose luminosity distribution parallels that of late-type
Irr/Pec systems, the analogues of today's dwarf galaxies
\citep{cross.bouwens.ea:luminosity,menanteau.ford.ea:morphological}.
\citet{kaviraj.khochfar.ea:uv} report a split in the color
distribution of $z\sim0.5$--1 E/S0s, with a ``minor but significant
peak'' on the blue sequence.  Several authors have suggested links
between blue compact systems and the bulges of today's large spirals
\citep{kobulnicky.zaritsky:chemical,barton.:possible,puech.hammer.ea:3d}.
However, both blue/blue-cored spheroids and blue compact galaxies are
composite populations, containing galaxies representing a wide range
of masses, environments, and formation scenarios
\citep[][]{lee.lee.ea:nature,
  noeske.koo.ea:luminous,barton..ea:search}.  Even at $z=0$, we have
seen that evolutionary scenarios for blue-sequence E/S0s are a
sensitive function of mass.  Also, the bulk of the systems we have
described reside in field environments, whereas blue or blue-cored
E/S0s in clusters are more likely to reflect transient states as a
prelude to quenching by harassment, stripping, or strangulation
\citep{koo.guzman.ea:on,rose.gaba.ea:starbursts,lisker.glatt.ea:virgo}.
Moreover, the abundance of AGNs among blue spheroids is disputed
\citep {lee.lee.ea:nature,treu.ellis.ea:assembly}.  Quantitative
assessment of the disk regrowth picture will therefore require careful
decomposition of high-$z$ samples to isolate analogues of the
disk-building blue-sequence E/S0s analyzed here.

\section{Conclusion}
\label{sec:concl}

We have examined the existence, properties, and evolutionary
trajectory of a substantial population of morphologically defined
E/S0s that fall on the blue sequence in color-stellar mass space,
i.e., the usual locus of late-type disk galaxies.  Our analysis of
blue-sequence E/S0s has relied on two independent data sets (the NFGS
and HyperLeda+ samples) and uses a third survey to calibrate
completeness (the NYU VAGC low-redshift sample).  Key results are
summarized here.

\begin{itemize}

\item{We identify three stellar mass scales of interest: the
  ``shutdown mass'' $\log{M_s/\msun}\sim11.2$ above which
  blue-sequence E/S0s (and the blue sequence) generally do not exist;
  the ``threshold mass'' $\log{M_t/\msun}\sim9.7$ below which
  blue-sequence E/S0s become extremely abundant; and the ``bimodality
  mass'' $\log{M_b/\msun}\sim10.5$ marking the midpoint of this range.
  Blue-sequence E/S0s increase in numbers down to $M_b$, then level
  off at 5--7\% of all E/S0s from $M_b$ to $M_t$, and finally spike up
  to $\sim$20--30\% of all E/S0s below $M_t$.  Our samples do not
  probe below $M_*\sim10^8$ \msun, and statistical analysis of
  demographics and environments is limited to $M_*>10^9$ \msun.}

\item{The bimodality mass $M_b$ and the threshold mass $M_t$ have been
  previously linked with strong shifts in star formation history and
  gas richness by \citet{kauffmann.heckman.ea:dependence} and
  \citet{kannappan:linking}, respectively (see also
  \citealt{kannappan.wei:galaxy}).  However, our definition of $M_t$
  corrects an erroneous association of $M_t$ with $M_b$ made by
  \citet{kannappan:linking}, caused by assuming agreement between the
  stellar mass zero points used by
  \citet{kauffmann.heckman.ea:dependence} and
  \citet{bell.mcintosh.ea:optical}. The stellar mass zero point used
  in the present work agrees in normalization with
  \citet{kauffmann.heckman.ea:dependence} and
  \citet{kannappan.gawiser:systematic}.}

\item{Morphologies and numbers of blue-sequence E/S0s with masses near
  $M_s$ are consistent with merger remnants that will fade up to the
  red sequence (``quenching mergers'').}

\item{Blue-sequence E/S0s below $M_b$ have fairly settled morphologies
  and numbers in the population exceeding the rate of close pairs,
  suggesting disk (re)building may be common.  Most are S0s (including
  S0/as), and all of the Es have some type of disk.}

\item{Below $M_b$, we measure abruptly lower density environments for
  both red- and blue-sequence E/S0s in our data sets, often even lower
  than for many late-type galaxies.  While these data are incomplete,
  we conclude that the E/S0s in our analysis represent a distinct
  population from dEs, which would be found primarily in a lower mass
  range in cluster environments.}

\item{Blue-sequence E/S0s are intermediate between red-sequence E/S0s
  and spirals in $M_*$-$\sigma$ and $M_*$-radius scaling relations,
  but are more similar to red-sequence E/S0s, validating their
  morphological classification.  Given that most sub-$M_b$ E/S0s in
  our samples reside in modest density environments, their compact
  structure is most naturally explained by past merging. We note the
  emergence of a shifted $M_*$-radius relation below $\sim$$M_t$.}

\item{Direct evidence for disk-building in blue-sequence E/S0s
  includes blue outer disk colors and a high frequency of
  kinematically distinct (polar or counterrotating) secondary stellar
  disks. Statistically, the strongest color difference between
  red- and blue-sequence E/S0s is in outer disk color.}

\item{Based on current gas content and specific star formation rates,
  the growth potential of blue-sequence E/S0s is comparable to that of
  late-type galaxies below $M_b$. Moreover, several lines of evidence
  suggest that star formation in blue-sequence E/S0s is intermittent,
  involving both minor and significant starbursts. We also find
  evidence that gas and stellar disk growth may be partly decoupled.}

\item{Pseudobulges may form in tandem with extended disks in many
  blue-sequence E/S0s.  All sub-$M_b$ blue-sequence E/S0s in the
  NFGS have modest central velocity dispersions, and about half show
  evidence of minor disturbances and centrally enhanced star
  formation, suggesting gas inflow events. However, these results are
  preliminary given the small sample size.}

\item{Red-sequence E/S0s may also participate in disk regrowth.  Below
  $M_b$, red-sequence E/S0s show substantial diversity in radii,
  central velocity dispersions, gas fractions, and specific star
  formation rates.  A subset of these systems could represent ``off''
  states in a bursty cycle of disk regrowth.}

\item{A significant fraction of blue-sequence E/S0s seem to occupy a
  ``sweet spot'' in parameter space, with masses and environments in a
  regime characterized by abundant gas, combined with concentration
  indices near the value identified by
  \citet{kauffmann.heckman.ea:gas} as optimal for peak star formation
  efficiency.  This raises the intriguing possibility that the
  processes that form blue-sequence E/S0s actually {\it enable}
  galaxies to build large spiral disks and escape from the dwarf
  irregular regime.  Based on downsizing considerations, we conjecture
  that some fraction of $z=0$ spirals above $M_b$ may have grown from
  analogous blue-sequence E/S0s more massive than those seen today.}

\end{itemize}

Several areas remain for future work.  Spatially resolved analysis of
gas reservoirs, star formation, and star formation histories is
underway to quantify disk and bulge growth in both red- and
blue-sequence E/S0s.  This analysis will assess E/S0 evolutionary
trajectories and examine the ratio of fading-merger vs.\ disk-building
systems as a function of mass. Evaluating the spiral-galaxy progenitor
scenario will require comparative analysis of high-redshift
blue-sequence E/S0s and presumed spiral descendents today.  A much
more extensive analysis of the internal dynamics of today's E/S0s with
masses below $M_b$ is also essential to establish the plausibility of
their building dominant dynamically cold components, i.e.,
pseudobulges and spiral disks.  If this picture proves correct, then
blue-sequence E/S0s may represent the first direct evidence for a
substantial and well-defined class of merger-formed galaxies caught in the
act of rebuilding disks, as predicted by hierarchical models of galaxy
formation.

\acknowledgements We thank Douglas Mar for helpful discussions.  The
anonymous referee prompted us to perform a completeness analysis of
the HyperLeda+ sample that proved highly informative.  SJK was
partially supported by an NSF Astronomy and Astrophysics Postdoctoral
Fellowship under award AST-0401547.  AJB was partially supported by a
Jansky Postdoctoral Fellowship from the National Radio Astronomy
Observatory, which is operated by Associated Universities, Inc., under
cooperative agreement with the National Science Foundation.  We
acknowledge use of the HyperLeda database (http://leda.univ-lyon1.fr).
We have also made use of data products from the Two Micron All Sky
Survey (2MASS), which is a joint project of the University of
Massachusetts and the Infrared Processing and Analysis
Center/California Institute of Technology, funded by the National
Aeronautics and Space Administration and the National Science
Foundation.  We acknowledge use of the Sloan Digital Sky Survey
(SDSS), for which funding has been provided by the Alfred P. Sloan
Foundation, the Participating Institutions, the National Aeronautics
and Space Administration, the National Science Foundation, the
U.S. Department of Energy, the Japanese Monbukagakusho, and the Max
Planck Society. The SDSS Web site is http://www.sdss.org/.  The SDSS
is managed by the Astrophysical Research Consortium (ARC) for the
Participating Institutions. The Participating Institutions are The
University of Chicago, Fermilab, the Institute for Advanced Study, the
Japan Participation Group, The Johns Hopkins University, the Korean
Scientist Group, Los Alamos National Laboratory, the
Max-Planck-Institute for Astronomy (MPIA), the Max-Planck-Institute
for Astrophysics (MPA), New Mexico State University, University of
Pittsburgh, University of Portsmouth, Princeton University, the United
States Naval Observatory, and the University of Washington.


\end{document}